\documentclass[aps,pra,nofootinbib,showpacs, twocolumn]{revtex4}
\usepackage{mathrsfs}
\usepackage{amsfonts}
\usepackage{amssymb}
\usepackage{amsmath}
\usepackage{epsf,graphics,graphicx}
\usepackage{relsize}
\usepackage[T1]{fontenc}
\usepackage{subfigure}

\def\rank{\mathrm{rank}}
\begin{document}
\title{Topological phases and multi-qubit entanglement}
\author{Markus Johansson$^1$}
\email{markus.johansson@kvac.uu.se}
\author{Marie Ericsson$^1$}
\email{marie.ericsson@kvac.uu.se}
\author{Kuldip Singh$^2$}
\email{sciks@nus.edu.sg}
\author{Erik Sj\"oqvist$^{1,2}$}
\email{erik.sjoqvist@kvac.uu.se}
\author{Mark S. Williamson$^{2}$}
\email{cqtmsw@nus.edu.sg}
\affiliation{$^1$Department of Quantum Chemistry, Uppsala University, Box 518,
Se-751 20 Uppsala, Sweden, EU}
\affiliation{$^2$Centre for Quantum Technologies, National University of Singapore,
3 Science Drive 2, 117543 Singapore, Singapore}
\date{\today}
\begin{abstract}
Global phase factors of topological origin, resulting from cyclic local $\rm{SU}$ evolution, called topological phases, were first described in [Phys. Rev. Lett.   {\bf 90}, 230403 (2003)], in the case of entangled qubit pairs.
In this paper we investigate topological phases in multi-qubit systems as the result of cyclic local $\rm{SU(2)}$ evolution. These phases originate from the topological structure of the local $\rm{SU(2)}$-orbits and are an attribute of most entangled multi-qubit systems. We discuss the relation between topological phases and SLOCC-invariant polynomials and give examples where topological phases appear. A general method to find the values of the topological phases in an $n$-qubit system is described and a complete list of these phases for up to seven qubits is given.

\end{abstract}
\maketitle
\section{Introduction}

The topological origin of global phase factors resulting from cyclic local $\rm{SU(2)}$ evolutions of entangled pairs of qubits was first noted by Milman and Mosseri \cite{milmos,milma}.
They showed that a global $\pi$-phase is related to the double connectedness of the local $\rm{SU(2)}$ orbit.
This topological phase for two qubits has since then been experimentally observed in the context of polarization and spatial mode transformations of a laser beam \cite{souz}, and in the context of NMR \cite{du}.
More recently, Oxman and Khoury \cite{oxma} showed that topological phases are a feature of two-qudit systems, for arbitrary dimension $d$, and gave an explicit description of the topological structure of the local ${\rm{SU}}(d)$ orbits.
The topological nature of the phase factors is related to entanglement in the sense that they are only present in two-qubit systems with nonzero concurrence, or more generally, nonzero determinant of the coefficient matrix \cite{co} for qudit pairs.

Characterizing multipartite entanglement is a challenging problem due to the rich structure of entanglement classes.
Different approaches include the classification of orbits in terms of their stabilizer groups \cite{lin,car,car2}, parametrization of orbit spaces by non-local parameters \cite{acin}, inter-convertibility under local operations \cite{dur,versa,car2,schli}, and the study of entanglement invariants \cite{woot,koff,kempe,grassl,sudden,luq,luk,luquew,dok,osta}.

The coarsest characterization of entanglement properties is equivalence under stochastic local operations and classical communication, SLOCC, i.e., inter-convertibility of quantum states under the widest possible group of local invertible operations.
The SLOCC entanglement classes have been found for up to four qubits \cite{dur,versa}, and the algebra of polynomial entanglement invariants have been described \cite{koff,luq}.

The finest characterization of entanglement properties, in the sense of discriminating entanglement types, is equivalence under local unitary operations \cite{sudden}. The entanglement classes and algebra of entanglement invariants have been described for up to four qubits \cite{sudden,luquew}.
Beyond four qubits there is only a partial understanding of entanglement classes and invariant algebras \cite{luk,osta,dok}.

Since entanglement properties are by definition invariant under local unitary transformations, each local unitary orbit corresponds to a characterization of entanglement properties.
The qualitative properties of the orbit, such as dimension and topological structure, thus correspond to a qualitative characterization of entanglement. For example, the local $\rm{SU(2)}$ orbit of two entangled qubits is doubly connected whereas the orbit of a product state is simply connected \cite{milma}. Also in the case of three qubits the qualitative changes in entanglement properties correspond to qualitative changes in the orbit, as shown by Carteret and Sudbery \cite{car}.

In this paper, we investigate topological phases in multi-qubit systems undergoing cyclic local $\rm{SU(2)}$ evolution and their relation to multipartite entanglement properties.
Topological phases are qualitative properties of the local unitary orbits and as such an attribute of multipartite entanglement.
A qualitative characterization of entanglement properties in terms of the set of topological phases is a classification intermediate in coarseness between local unitary equivalence and SLOCC-equivalence.
We discuss the relation between topological phases and nonzero polynomial entanglement invariants.
Furthermore, we describe a method for finding the topological phases of an $n$-qubit system and apply this method explicitly for up to seven qubits.

The outline of the paper is as follows.
In Sec. \ref{udnurt}, we discuss the concept of topological phases and its relation to stabilizer groups.
We describe topological phases in the context of multi-qubit systems undergoing local $\rm{SU(2)}$ evolution in Sec. \ref{sec3}. We study some general existence criteria as well as give examples of states with topological phases. Section \ref{ent} contains a discussion of the relation between entanglement properties of a state and topological phases.
In particular, we focus on entanglement invariants and the special significance of maximally entangled states. In Sec. \ref{sec5}, we
introduce a method for finding the topological phases based on Cartan subgroups of the local $\rm{SU(2)}$ operations, including a combinatorial formulation of the problem.
Finally, in Sec. \ref{ytrrium}, we present a complete list of the topological phases that exist in $n$-qubit systems for $n$ up to $7$. These were found through a search algorithm based on the method introduced in Sec. \ref{ent}.
From the results we identify some SLOCC-classes that are distinguished by the different topological phases of their respective maximally entangled states. The paper end with the conclusions.

\section{Topological phases}

A \label{udnurt}quantum system, initially in pure state $|{\psi}\rangle$, undergoing evolution through a continuous closed path in its projective Hilbert space by action of a Lie-group $G$, acquires a  global phase factor $e^{i\chi}$, where $\chi\in\mathbb{C}$.
We allow $\chi$ to be a complex number to include the most general case of dissipative evolution and refer to it, in this case, as a complex phase.
The phase factor is determined by the $G$-action and the state. Let us now consider the case where $|{\psi}\rangle$ and $G$ are such that only a discrete set of phase factors are possible as the result of closed path evolution.
Then two closed paths associated with different phase factors cannot be continuously deformed into each other since this would require the phase factor to change continuously through the forbidden intermediate values.
Two such closed paths hence belong to different homotopy classes with $|{\psi}\rangle$ as the base point, on the $G$-orbit over $|{\psi}\rangle$.
The $G$-orbit over $|{\psi}\rangle$ is the immersed submanifold $\mathcal{O}^{G}_{|{\psi}\rangle}=\{g|{\psi}\rangle\mid{g\in{G},\phantom{u}|{\psi}\rangle\sim\lambda|{\psi}\rangle,\phantom{u}\lambda\in{\rm{\mathbb{C}\backslash\{0\}}}}\}$ of projective Hilbert space.
The discreteness of the set of possible phase factors thus implies that the topology of $\mathcal{O}^{G}_{|{\psi}\rangle}$ is non-trivial in the sense that it is not simply connected. We therefore denote the phases associated with closed path evolutions of these kind of states as {\it topological phases}.

The set of phase factors $e^{i\chi}$ associated with cyclic $G$-evolutions  of a state $|{\psi}\rangle$ are the solutions of the equations of the form

\begin{eqnarray}\label{eq0}g|\psi\rangle=e^{i\chi}|\psi\rangle,\phantom{uu}g\in{G},\phantom{uu}\chi\in{\mathbb{C}}.\end{eqnarray}
The phase factors $e^{i\chi}$ and the associated cyclic evolutions are closely related to stabilizers of the state $|{\psi}\rangle$ in the sense that if $g_{\chi}\in{G}$ is a group element giving the topological phase $\chi$, i.e., $g_{\chi}|{\psi}\rangle=e^{i\chi}|{\psi}\rangle$, and $\{g_{\chi}\}$ the set of all such elements, then the union of all such sets $\bigcup_{\forall{\chi}}\{g_{\chi}\}$ is the stabilizer subgroup of $G$ of the state $|{\psi}\rangle$, viewed as an element of projective Hilbert space. Likewise the set $\bigcup_{\forall{\chi}}\{e^{-i\chi}g_{\chi}\in{\rm{\mathbb{C}\backslash\{0\}}\otimes{G}}\}$, is the stabilizer subgroup of $\rm{\mathbb{C}\backslash\{0\}}\otimes{G}$ of the state $|{\psi}\rangle$, when viewed as an element of Hilbert space.

The set of $g_{\chi}$ for each $\chi$ can be a discrete set of group elements, or a discrete set of continuous families $g_{\chi}(\alpha_{1},\dots,\alpha_{n})$ parametrized by real parameters $\alpha_{1},\dots,\alpha_{n}$.
For each family $g_{\chi}(\alpha_{1},\dots,\alpha_{n})$, possibly a family with only one element, there is a homotopy class of curves, parametrized by $t,\alpha_{1},\dots,\alpha_{n}$, given by $\{g(t), t\in[0,1]\mid{g}(0)=\hat{1}, g(1)\in{g_{\chi}(\alpha_{1},\dots,\alpha_{n})}\}$ in $G$. Each such class, when acting on $|{\psi}\rangle$, defines a homotopy class of cyclic evolutions with $|{\psi}\rangle$ as base point, in $\mathcal{O}^{G}_{|{\psi}\rangle}$, corresponding to the phase $\chi$. We can therefore use the families $g_{\chi}(\alpha_{1},\dots,\alpha_{n})$ to represent the corresponding homotopy classes.
Furthermore, the phase factor associated with the class of contractible loops is by necessity $+1$. Therefore the choice of base point for the closed paths in $\mathcal{O}^{G}_{|{\psi}\rangle}$ does not alter the set of available topological phases.

\section{Topological phases in multi-qubit systems}

In \label{sec3}the following, we consider the case of topological phases of a system of $n$ qubits acted upon by the group of $n$-local special unitary operations ${\rm{SU(2)}}^{\otimes{n}}$. We thus consider equations of the form

\begin{eqnarray}\label{eq1}U|\psi\rangle=e^{i\chi}|\psi\rangle,\phantom{uu}U\in{{\rm{SU(2)}}^{\otimes{n}}},\phantom{uu}\chi\in\mathbb{R}.\end{eqnarray}
Note that since we are considering non-dissipative evolution $\chi$ is a real number. The restriction to local special unitary operations, rather than the full group of local unitary operations, is made since the set of phase factors that results from cyclic ${\rm{U(2)}}^{\otimes{n}}$ evolution is always continuous, and would therefore not reveal any information about the topology of $\mathcal{O}^{{\rm{U(2)}}^{\otimes{n}}}_{|{\psi}\rangle}$.

To elucidate the question of appearance of topological phases in $n$-qubit systems we make the following observation.
A criterion for existence of topological phases, satisfied for a wide class of states, follows immediately from the theory of polynomial invariants \cite{mum,hilb}.
Let us consider the states that can be brought to a form where every reduced one-qubit density matrix is maximally mixed by a sequence (possibly infinite) of ${\rm{SL(2,\mathbb{C})}}^{\otimes{n}}$-operations.
Such states belong to a class for which there exist a nonzero polynomial in the expansion coefficients of the state, when represented in a local basis, that is invariant under ${\rm{SL(2,\mathbb{C})}}^{\otimes{n}}$-operations \cite{vers}. The existence of such an invariant puts a restriction on the set of phases $\chi$ that can result from cyclic evolutions, since a polynomial of degree $d$ in the expansion coefficients can only be invariant under multiplication by phase factors $e^{i\chi}$ satisfying $(e^{i\chi})^{d}=1$. The set of $\chi$ is discrete and are therefore topological phases.

The property of a state of being related to a state with maximally mixed reduced one-qubit density matrices by a sequence (possibly infinite) of ${\rm{SL(2,\mathbb{C})}}^{\otimes{n}}$-operations means that the state is ${\rm{SL(2,\mathbb{C})}}^{\otimes{n}}$-semistable \cite{mum,vers,dolg}.
A state $|\psi\rangle$ is ${\rm{SL(2,\mathbb{C})}}^{\otimes{n}}$-semistable if the $0$-vector is not included in the closure $\overline{\mathcal{O}_{|{\psi}\rangle}^{{\rm{SL(2,\mathbb{C})}}^{\otimes{n}}}}$ of the ${\rm{SL(2,\mathbb{C})}}^{\otimes{n}}$-orbit. In other words, the norm of $|\psi\rangle$ cannot approach zero asymptotically under action of ${\rm{SL(2,\mathbb{C})}}^{\otimes{n}}$. It is this property that implies that $|\psi\rangle$ can be separated from 0 by some ${\rm{SL(2,\mathbb{C})}}^{\otimes{n}}$ polynomial invariant $I$, that is $I(|{\psi}\rangle)\neq{I(0)}$ and $I(|{\psi}\rangle)={I(g|{\psi}\rangle)}$ for $g\in{\rm{SL(2,\mathbb{C})}}^{\otimes{n}}$ \cite{vers}. States that are not semistable are termed ${\rm{SL(2,\mathbb{C})}}^{\otimes{n}}$-unstable. The relation between topological phases, entanglement properties, and polynomial invariants is discussed in greater detail in Sec. \ref{ent}.

From the above we can conclude that all ${\rm{SL(2,\mathbb{C})}}^{\otimes{n}}$-semistable states have topological phases. Furthermore, some of the ${\rm{SL(2,\mathbb{C})}}^{\otimes{n}}$-unstable states also have topological phases. To illustrate the different cases we consider some examples of ${\rm{SL(2,\mathbb{C})}}^{\otimes{n}}$-semistable and -unstable states. Here and throughout the rest of the paper we will ignore the normalization of states.

An example of an $\rm{SL(2,\mathbb{C})^{\otimes{3}}}$-semistable three-qubit state is the GHZ-state $|\rm{GHZ}\rangle=|000\rangle+|111\rangle$. The cyclic $\rm{SU(2)^{\otimes{3}}}$ evolutions correspond to two different sets of families of stabilizers \cite{car}. One set, containing two families distinguished by integers $p=0,1$, where each family is  parametrized by real numbers $\alpha$ and $\beta$, is

\begin{eqnarray}U(p,\alpha,\beta)=e^{i\alpha\sigma_{z}^{1}}\otimes{e}^{i\beta\sigma_{z}^{2}}\otimes{e}^{i(p\pi-\alpha-\beta)\sigma_{z}^{3}},\end{eqnarray}
where ${\sigma}_{z}^{k}$ is the ${\sigma}_{z}$ operation on the $k$th qubit. The corresponding topological phases are $\chi=0$ for $p=0$ and $\chi=\pi$ for $p=1$.
The other set, containing two families distinguished by integers $q=0,1$, where each family is parametrized by two real numbers $\gamma$ and $\delta$, is

\begin{eqnarray}\label{crit}U(q,\gamma,\delta)=\left(\!\!\begin{array}{cc}
0 & e^{i\gamma} \\
-e^{-i\gamma} & 0\\
\end{array}\!\!\right)\!\!\otimes\!\!\left(\!\!\begin{array}{cc}
0 & e^{i\delta} \\
-e^{-i\delta} & 0\\
\end{array}\!\!\right)\!\!\otimes\nonumber\\
\!\!\left(\!\!\begin{array}{cc}
0 & e^{i(\frac{\pi}{2}+q\pi-\gamma-\delta)} \\
-e^{-i(\frac{\pi}{2}+q\pi-\gamma-\delta)} & 0\\
\end{array}\!\!\right).\end{eqnarray}
The topological phases corresponding to these families are $\chi=\frac{\pi}{2}$ for $q=0$ and $\chi=-\frac{\pi}{2}$ for $q=1$. The four different families of these two sets correspond to four different homotopy classes (of which one is trivial) of cyclic evolutions on $\mathcal{O}_{|{\rm{GHZ}}\rangle}^{{\rm{SU(2)}}^{\otimes{3}}}$.

The generalized $n$-qubit GHZ-states $|{\rm{GHZ}}^{n}\rangle=\bigotimes_{k=1}^{n}|{0_{k}}\rangle+\bigotimes_{k=1}^{n}|{1_{k}}\rangle$ have families of stabilizers parametrized by $n-1$ real parameters and integers $0,1$. These are a direct generalization of the stabilizers of the three-qubit GHZ-state, and read

\begin{eqnarray}U(\alpha_{1},\dots,\alpha_{(n-1)})=\bigotimes_{k=1}^{n}e^{i\alpha_{k}\sigma_{z}^{k}},\end{eqnarray}
where $\sum_{k=1}^{n}\alpha_{k}=p\pi$, $p=0,1$, for which $\chi=0,\pi$, and

\begin{eqnarray}\label{66}U(\delta_{1},\dots,\delta_{(n-1)})=\bigotimes_{k=1}^{n}\left(\!\!\begin{array}{cc}
0 & e^{i\delta_{k}} \\
-e^{-i\delta_{k}} & 0\\
\end{array}\!\!\right),\end{eqnarray}
where $\sum_{k=1}^{n}\delta_{k}=\frac{\pi}{2}+q\pi$, $q=0,1$, for odd $n$ and $\sum_{k=1}^{n}\delta_{k}=+q\pi$, $q=0,1$, for even $n$. The topological phases associated with the families in Eq.~(\ref{66}) are $\chi=\pm\frac{\pi}{2}$ for odd number of qubits and $\chi=0,\pi$ for even number of qubits.

Another class of $n$-qubit states, that coincide with the GHZ-state for three qubits, but where the topological phases have a different $n$-dependence, are the states of the form
\begin{eqnarray}\label{yhn}|\psi\rangle=\bigotimes_{k=1}^{n}\!|{1_{k}}\rangle+\sum_{r=1}^{n}\!|{1_{r}}\rangle\bigotimes_{k\neq{r}}^{n}\!|{0_{k}}\rangle\equiv{\bigotimes_{k=1}^{n}\!|{1_{k}}\rangle+|{W}^n\rangle},\nonumber\\
\end{eqnarray}
for which the stabilizers, parametrized by integers $q_{k}$, are
\begin{eqnarray}\label{trys}U(q_{0},q_{1},\dots,q_{n})=\bigotimes_{k=1}^{n}e^{i\left[\frac{\mathlarger{\left({q_{0}-\sum_{j=1}^{n}q_{j}}\right)}}{\mathlarger{\left(n-1\right)}}+q_{k}\right]\pi\sigma_{z}^{k}},\end{eqnarray}
and the corresponding topological phases are $\chi=\frac{\sum_{k=0}^{n}q_{k}\pi}{(n-1)}$.

We next turn to the ${\rm{SL(2,\mathbb{C})}}^{\otimes{n}}$-unstable states. These are the complement of the semistable states in Hilbert space, and contain all the states that do not have topological phases. However, as previously mentioned, there are states in this category that do have topological phases.
A general result concerning which properties of an ${\rm{SL(2,\mathbb{C})}}^{\otimes{n}}$-unstable state that implies existence of topological phases is beyond the scope of the present study. Instead, we consider a few examples of ${\rm{SL(2,\mathbb{C})}}^{\otimes{n}}$-unstable states to demonstrate that there are states in this category with topological phases for any number of qubits, as well as states without topological phases for any number of qubits.

For three qubits we can consider the $W$-state $|W\rangle=|100\rangle+|010\rangle+|001\rangle$. In any basis this state can be expressed on the form

\begin{eqnarray}|W\rangle=|1_{1}\rangle|0_{2}\rangle|0_{3}\rangle+|0_{1}\rangle(|0_{2}\rangle|1_{3}\rangle+|1_{2}\rangle|0_{3}\rangle),\end{eqnarray}
where $|0_{k}\rangle$ and $|1_{k}\rangle$ are orthogonal states of the $k$th qubit.
Since qubits $2$ and $3$ are in a product state in the first term and in a maximally entangled state in the second term, these two terms cannot be brought to each other by $\rm{SU(2)^{\otimes{3}}}$ operations. Therefore, in a cyclic evolution each of the two terms must evolve back into themselves.
Furthermore, focusing on the first qubit, the only $\rm{SU(2)}$ operations for which both $|1_{1}\rangle$ and $|0_{1}\rangle$ are eigenvectors, are the diagonal operators $e^{i\alpha{\sigma}_{z}^{1}}$. Due to permutation symmetry we can conclude that each local $\rm{SU(2)}$ operation has to be on the same form, up to an overall sign. It follows that the full group of stabilizers corresponding to cyclic evolutions of the $W$-state is

 \begin{eqnarray}U(\alpha,q_{1},q_{2},q_{3})\!=e^{i(\alpha+q_{1}\pi){\sigma}_{z}^{1}}\!\!\otimes\!{e}^{i(\alpha+q_{2}\pi){\sigma}_{z}^{2}}\!\!\otimes\!{e}^{i(\alpha+q_{3}\pi){\sigma}_{z}^{3}}\nonumber\\
 \end{eqnarray}
with $\alpha$ an arbitrary real number and $q_{1},q_{2},q_{3}$ integers. From this we see that the global phase $\chi=\alpha\pm\pi$ can take any value. The $W$-state is thus an example of an ${\rm{SL(2,\mathbb{C})}}^{\otimes{3}}$-unstable state that does not have topological phases.

The generalized $n$-qubit $W$-state $|{W}^n\rangle=\sum_{r=1}^{n}|{1_{r}}\rangle\bigotimes_{k\neq{r}}^{n}|{0_{k}}\rangle$, with stabilizers $\bigotimes_{k=1}^{n}e^{i(\alpha+q_{k}\pi){\sigma}_{z}^{k}}$, does not have topological phases either.
We can thus conclude that there will be entangled states without topological phases for arbitrary number of qubits.

Having seen this, let us now instead consider examples of $\rm{SL(2,\mathbb{C})^{\otimes{n}}}$-unstable states with topological phases.
For three qubits, we consider the state $|\psi\rangle=|000\rangle+|100\rangle+|010\rangle+|001\rangle=|000\rangle+|{W}\rangle$, that belong to the $W$ SLOCC-class \cite{dur}. In any basis, this state can be expressed as
\begin{eqnarray}\label{tewf}|\psi\rangle&=&(|0_{1}\rangle+|1_{1}\rangle)|0_{2}\rangle|0_{3}\rangle\nonumber\\
&&+|0_{1}\rangle(|0_{2}\rangle|1_{3}\rangle+|1_{2}\rangle|0_{3}\rangle),\nonumber\\
\end{eqnarray}
where again $|0_{k}\rangle$ and $|1_{k}\rangle$ are orthogonal states of the $k$th qubit. Also in this case the two terms must evolve into themselves under a cyclic $\rm{SU(2)^{\otimes{3}}}$ evolution. The difference compared to the $W$-state is that, considering again the first qubit, the only $\rm{SU(2)}$ operations, for which both $|0_{1}\rangle+|1_{1}\rangle$ and $|0_{1}\rangle$ are eigenvectors, are $\pm{\hat{1}_{1}}$. Due to permutation symmetry it can be seen that the only cyclic evolutions of $|\psi\rangle$ are the ones corresponding to the stabilizers $\pm{\hat{1}_{1}}\otimes\pm{\hat{1}_{2}}\otimes\pm{\hat{1}_{3}}$, and that $\chi=\pm{\pi}$. Thus, $|000\rangle+|{W}\rangle$ has topological phases.

Generalizing this construction to $n$-qubits we construct the states

\begin{eqnarray}\label{yhn2}|\psi\rangle=\bigotimes_{k=1}^{n}|0_{k}\rangle+\sum_{r=1}^{n}|{1_{r}}\rangle\bigotimes_{k\neq{r}}^{n}|{0_{k}}\rangle\equiv{\bigotimes_{k=1}^{n}|{0_{k}}\rangle+|{W}^n\rangle},\nonumber\\
\end{eqnarray}
The cyclic evolutions correspond to the stabilizers $\bigotimes_{k=1}^{n}e^{i\alpha{\sigma}_{z}^{k}}$ where $\alpha=0,\pi$, and $\chi=0,\pi$ for all $n$.

Finally, we make a few observations about the stabilizers corresponding to cyclic evolutions associated with a topological phase $\chi$. Two stabilizer subgroups associated with two different states that belong to the same local unitary orbit are conjugated. More precisely, for a stabilizer $U$ of a state $|\psi\rangle$, there will be for every local unitary operator $V$, a conjugated stabilizer $VUV^{\dagger}$ of the state $V|\psi\rangle$ associated with the same topological phase.
Furthermore, the stabilizer $U$ is a stabilizer of all states related to $|\psi\rangle$ by a local filtering operation $D$ (a diagonalizable SLOCC-operation) that diagonalize in the same basis as $U$. The conjugated stabilizer $VUV^{\dagger}$ is likewise a stabilizer of any state related to $V|\psi\rangle$ by the local filtering operation $VDV^{\dagger}$. A local unitary operation $V$ on $|\psi\rangle$ followed by a local filtering operations in the basis for which $VUV^{\dagger}$ diagonalize is equivalent to a local filtering operation in the basis for which $U$ diagonalize followed by a unitary operation $V$ since $VD=VDV^{\dagger}V$. From this we can conclude that all states related to $|\psi\rangle$ by the set of SLOCC-operations on the form $VD$ and their local unitary orbits all share the topological phase $\chi$. This equivalence class of states with a common stabilizer, up to conjugation, associated with a topological phase $\chi$ is thus wider than a unitary equivale
 nce class but not a full SLOCC-class. We call this family of unitary orbits the {\it $A$-class associated with $U$}. The $A$-equivalence constitute a classification of entanglement intermediate in coarseness between SLOCC-equivalence and local unitary equivalence.

\section{Relation between topological phases and entanglement}

As already seen there is a relation between the topological phases of a state and its entanglement properties.
We \label{ent}now examine this relation in some more detail.

\begin{figure}[h]
\includegraphics[width=8.0 cm]{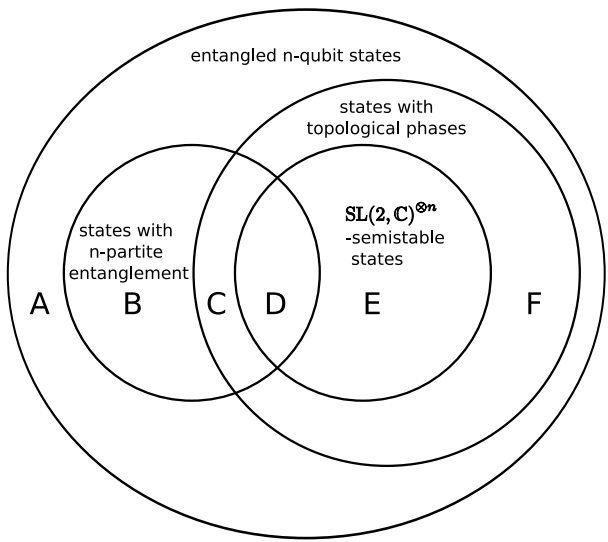}
\caption{\footnotesize{Venn diagram of the set of entangled $n$-qubit states with the subsets; states with $n$-partite entanglement (i.e. inseparable over every bipartition of the qubits), ${\rm{SL(2,\mathbb{C})}}^{\otimes{n}}$-semistable states, and states with topological phases. The six nonempty intersections of the sets and the complements of the sets are labeled A,B,C,D,E, and F, and contain the following classes of states. A: Entangled states separable over at least one bipartition and with no topological phases. Example $|{{W}}^{n-m}\rangle\otimes|{{W}}^{m}\rangle$. B: States with $n$-partite entanglement that are ${\rm{SL(2,\mathbb{C})}}^{\otimes{n}}$-unstable and have no topological phases. Example $|W^{n}\rangle$. C: States with $n$-partite entanglement that are ${\rm{SL(2,\mathbb{C})}}^{\otimes{n}}$-unstable and have topological phases. Example $|0\dots0\rangle+|W^{n}\rangle$. D: States with $n$-partite entanglement that are ${\rm{SL(2,\mathbb{C})}}^{\otimes{n}}$-s
 emistable. Example $|{\rm{GHZ}}^{n}\rangle$. E: ${\rm{SL(2,\mathbb{C})}}^{\otimes{n}}$-semistable states that are separable over at least one bipartition. Example $|{\rm{GHZ}}^{n-m}\rangle\otimes|{\rm{GHZ}}^{m}\rangle$. F: ${\rm{SL(2,\mathbb{C})}}^{\otimes{n}}$-unstable states separable over at least one bipartition with topological phases. Example $|{\rm{GHZ}}^{n-m}\rangle\otimes(|0\dots0\rangle+|{{W}}^{m}\rangle)$.}}
\label{fig}
\end{figure}

Through examples we have established that for arbitrary number of qubits there is a hierarchy of states, where the ${\rm{SL(2,\mathbb{C})}}^{\otimes{n}}$-semistable states constitute the smallest set, contained within the larger set of states with topological phases, being in turn a subset of the full Hilbert space. All of these examples are states with $n$-partite entanglement, i.e., states that are inseparable over any bipartition of the qubits. Furthermore, no state where at least one qubit is in a tensor product with the rest of the qubits, can have a topological phase. This follows from the fact that a single qubit in a pure state can be given an arbitrary phase factor by closed path $\rm{SU(2)}$ evolution.
A state of the form $|{\psi}\rangle=|{\varrho}\rangle\otimes|{\theta\rangle}$ has a topological phase $\chi$ if and only if $|{\varrho}\rangle$ and $|{\theta\rangle}$ have topological phases $\chi_{\varrho}$ and $\chi_{\theta}$, respectively. In such a case $\chi=\chi_{\varrho}+\chi_{\theta}$. The relation between the set of states with topological phases, the set of states with $n$-partite entanglement, and the set of ${\rm{SL(2,\mathbb{C})}}^{\otimes{n}}$-semistable states is illustrated in Fig. \ref{fig}.

Any ${\rm{SL(2,\mathbb{C})}}^{\otimes{n}}$-invariant polynomial is invariant only if the state is not normalized after an ${\rm{SL(2,\mathbb{C})}}^{\otimes{n}}$-operation. If the resulting state is normalized, the polynomial instead becomes a homogenous function of the coefficients $b_{j}$, that has its maximum on the set of states for which all reduced rank-$2$ density matrices are maximally mixed \cite{vers}. Therefore, arguing that these kind of homogenous functions are measures of entanglement, it was suggested that the states for which all reduced rank-$2$ density matrices are maximally mixed, should be viewed as the maximally entangled states within their respective SLOCC-orbit. The maximally entangled state of a SLOCC-orbit is unique up to local unitary transformations \cite{vers,kempf}. Thus, there is at most one local unitary orbit that is maximally entangled in each SLOCC-orbit.

From the point of view of topological phases these maximally entangled states have a special significance in relation to the geometric phase \cite{aharanovanandan}. Generally, the topological phase consists of a dynamical and a geometric phase. For the maximally entangled states, the dynamical phase is zero and thus the acquired topological phase $\chi$ is identical to the geometric phase. The geometric phase for a cyclic evolution of a maximally entangled state is therefore of topological origin.

The ${\rm{SL(2,\mathbb{C})}}^{\otimes{n}}$-unstable states that do have topological phases, but for which there are no nonzero ${\rm{SL(2,\mathbb{C})}}^{\otimes{n}}$-invariants, cannot be brought to a form with maximally mixed reduced rank-$2$ density matrices. Therefore, there are no such states for which the acquired phase is purely geometric for all cyclic evolutions belonging to the same homotopy class.

As already discussed in Sec. \ref{sec3} there is a relation between polynomial ${\rm{SL(2,\mathbb{C})}}^{\otimes{n}}$-invariants and topological phases in the sense that knowledge of an invariant imposes restrictions on the possible values of topological phases, and vice versa, knowledge of a topological phase imposes a restriction on the form of polynomial invariants. Now, let $\chi_{min}$ be the smallest possible topological phase that a specific state can attain as the result of cyclic evolution. For convenience we chose $\chi$ to be positive.
The existence of an ${\rm{SL(2,\mathbb{C})}}^{\otimes{n}}$-invariant polynomial of degree $d$ implies that the states in the corresponding SLOCC-class where the invariant is nonzero can
only have $\chi_{min}$ that are multiples of $\frac{2\pi}{d}$. The value $\frac{2\pi}{d}$ of $\chi_{min}$ is the smallest possible given the degree of the polynomial, but this value need not be realized by any cyclic evolution. If several ${\rm{SL(2,\mathbb{C})}}^{\otimes{n}}$-invariant polynomials of degrees $d_{1},\dots,d_{k}$ are nonzero for some SLOCC-class, this limits the possible values of $\chi_{min}$ for this class to be a multiple of $\frac{2\pi}{gcd(d_{1},\dots,d_{k})}$, where $gcd(d_{1},\dots,d_{k})$ is the greatest common divisor of $\{d_{1},\dots,d_{k}\}$. Vice versa, the existence of an ${\rm{SL(2,\mathbb{C})}}^{\otimes{n}}$-semistable state such that $\chi_{min}=\frac{2\pi}{p}$, where $p$ is an integer, implies that there must be an ${\rm{SL(2,\mathbb{C})}}^{\otimes{n}}$-invariant polynomial of even degree $d$ such that $\frac{d}{p}$ is an integer. The degree $d$ must be even since no nonzero ${\rm{SL(2,\mathbb{C})}}^{\otimes{n}}$-invariant polynomial of odd d
 egree exist.

The algebra of polynomial ${\rm{SL(2,\mathbb{C})}}^{\otimes{n}}$-invariants is finitely generated, i.e., there exist a finite number of polynomials that generate the algebra \cite{hilb2}. Since every monomial part of an invariant must contain at least one of the generators as a factor, it follows that for every state for which a polynomial invariant is nonzero, one of the generators must be nonzero as well. This means that the degree of the generating polynomials puts the ultimate limit on what topological phases can exist in a system of qubits. If the degree of the generators are $d_{1},d_{2},\dots,d_{k},$ this restricts the set of topological phases to be a subset of  $\frac{2m_{1}\pi}{d_{1}},\frac{2m_{2}\pi}{d_{2}},\dots,\frac{2m_{k}\pi}{d_{k}}$, where $m_{k}$ are integers.

For two qubits the well known $\pi$-phase \cite{milmos,milma} is the only possible topological phase.
This can be understood from the fact that the only generator of the $\rm{SL(2,\mathbb{C})^{\otimes{2}}}$-invariant polynomials for two qubits is the complex concurrence, or preconcurrence ${\mathcal{C}}$ \cite{woot} which is a polynomial of degree $2$, related to
the concurrence $C$ by $C=|\mathcal{C}|$.

For three qubits in addition to the $\pi$-phase there is a $\frac{\pi}{2}$-phase. This phase is associated with the GHZ-states. Since the corresponding $\rm{SL(2,\mathbb{C})^{\otimes{3}}}$-invariant polynomial, the complex $3$-tangle $\mathcal{\scriptstyle{T}}_{3}$ (related to the 3-tangle $\tau_{3}$ \cite{koff} as $\tau_{3}=|\mathcal{\scriptstyle{T}}_{3}|$) is a polynomial of degree four, it is clear that no states in the GHZ SLOCC-class could have an associated $\chi_{min}$ other than $\pi$ and $\frac{\pi}{2}$. The family of stabilizers in Eq.~(\ref{crit}) of the GHZ-state correspond to the $\frac{\pi}{2}$ phase, and each element of this family is a stabilizer of its associated $A$-class. The union of all such orbits is however not the full GHZ SLOCC-class. We therefore see that the $A$-classes associated with the stabilizers in Eq.~(\ref{crit}) of the GHZ-state, which is the maximally entangled state of the GHZ SLOCC-class, is distinguished by a larger set of topological p
 hases.

For four qubits it was shown by Luque and Thibon that the generating invariant polynomials are of degree $2$, $4$, and $6$ \cite{luq}. Therefore the only possible $\chi_{min}$ for the SLOCC-classes, for which these generators are nonzero, are $\pi$, $\frac{\pi}{2}$, and $\frac{\pi}{3}$ respectively.
The state in Eq.~(\ref{yhn}) for $n=4$ is a state for which $\chi_{min}=\frac{\pi}{3}$ and thus such a phase exists.

The states of the form in Eq.~(\ref{yhn}) belong to the generic family $G_{abcd}$, parametrized by four complex numbers, of SLOCC-classes in the Verstraete {\it{et al.}} \cite{versa} classification and this is the only family containing states with $\chi_{min}=\frac{\pi}{3}$. It should be noted that the SLOCC-classes containing elements of the form in Eq.~(\ref{yhn}) belong to subfamilies of $G_{abcd}$ characterized by only one complex number. In contrast, the states with topological phase $\frac{\pi}{2}$ in $G_{abcd}$ belong to subfamilies characterized by two complex numbers. Furthermore, all states in $G_{abcd}$ have the topological phase $\pi$. From this we can see that a randomly chosen state in $G_{abcd}$ will, with probability one, have only the topological phase $\pi$.

Going in the other direction we can say that the existence of a topological phase $\frac{\pi}{d}$ implies that there must be a generator polynomial of degree $2dm$ where $m$ is an integer. More precisely, if a state has a topological phase $\frac{\pi}{d}$, the only polynomials that are nonzero on the ${\rm{SU(2)}}^{\otimes{n}}$-orbit of the state are of degrees $2dm$ and at least one of those must be a generator.

So far we have focused only on ${\rm{SL(2,\mathbb{C})}}^{\otimes{n}}$-invariant polynomials in the coefficients of the state. We may also consider the wider class of ${\rm{SU(2)}}^{\otimes{n}}$-invariant polynomials \cite{luquew} in the coefficients $b_{j}$ and the complex conjugated coefficients $b_{j}^*$. If such a polynomial is of degree $d_{1}$ in $b_{j}$ and degree $d_{2}\neq{d_{1}}$ in $b_{j}^*$, and is nonzero on an $\rm{SU(2)^{\otimes{n}}}$-orbit, this orbit has a discrete set of phases $\chi$ associated with the set of cyclic evolutions and the restriction on $\chi$ is that $e^{i(d_{1}-d_{2})\chi}=1$. Hence a non-trivial topological phase must satisfy $\chi=\frac{2m\pi}{d_{1}-d_{2}}$.

Furthermore, as pointed out previously, the eigenvalue problem in Eq.~(\ref{eq1}) for the unitary group ${\rm{U(2)}}^{\otimes{n}}$ do not give us any information about the topology of the orbits. The existence of a nonzero polynomial ${\rm{U(2)}}^{\otimes{n}}$-invariant does not give any restriction on the set of phases that are solutions of Eq.~(\ref{eq1}), and thus gives no information about the topological structure either. This is because a ${\rm{U(2)}}^{\otimes{n}}$-invariant polynomial is invariant under multiplication of the state by any global phase, which means it is of equal degree in $b_{j}$ and $b_{j}^*$.

The set of topological phases associated with an ${\rm{SU(2)}}^{\otimes{n}}$-orbit, or with an ${\rm{SL(2,\mathbb{C})}}^{\otimes{n}}$-orbit, is obviously a property that is invariant under $\rm{SU(2)^{\otimes{n}}}$ and ${\rm{SL(2,\mathbb{C})}}^{\otimes{n}}$-operations, respectively. As such they can be viewed as non-polynomial entanglement invariants, that give a qualitative characterization of the entanglement properties of the state.

\section{Finding topological phases for n qubits}

\label{sec5}
\subsection{Cartan subgroups and balanced states}

\label{sec52}
Having discussed the existence of topological phases, and their relation to entanglement properties, we now turn to the question of calculating the values of $\chi$ given a state and finding the full set of topological phases $\chi$ that can exist in an $n$-qubit system.

A way to approach the problem of finding the set of $e^{i\chi}$ associated with closed paths on an ${\rm{SU(2)}}^{\otimes{n}}$-orbit of a state $|{\psi}\rangle$ is to consider Eq.~(\ref{eq1}) for each  maximal Abelian subgroup of ${\rm{SU(2)}}^{\otimes{n}}$. A maximal Abelian subgroup of ${\rm{SU(2)}}^{\otimes{n}}$, also called a Cartan subgroup, is the set of all ${\rm{SU(2)}}^{\otimes{n}}$-operators that diagonalize in the same basis.
Every element in ${\rm{SU(2)}}^{\otimes{n}}$ is diagonalizable, i.e., it belongs to some Cartan subgroup.
Therefore, finding for each Cartan subgroup, the set of $e^{i\chi}$ associated with the cyclic evolutions with corresponding stabilizers in this Cartan subgroup, and combining the sets of $e^{i\chi}$ corresponding to the different Cartan subgroups, gives the complete set for ${\rm{SU(2)}}^{\otimes{n}}$. Thus, a state has topological phases if and only if the set of phase factors is discrete for every Cartan subgroup.
This is the method we will use to find the topological phases. To see how the method works, we discuss some properties of states when acted on by Cartan subgroups and demonstrate that it is sufficient to consider a special class of states in order to find all topological phases.

All operators in a Cartan subgroup have the same set of eigenvectors, namely the vectors corresponding to any basis in which the Cartan group diagonalize. We can always choose these to be a basis of product state vectors.

To describe such a basis we let $U=\bigotimes_{k=1}^{n}U_{k}$ be an ${\rm{SU(2)}}^{\otimes{n}}$-operator in the Cartan subgroup, where $U_{k}$ is the local $\rm{SU(2)}$-operator acting on the $k$th qubit. The eigenvectors of $U$ are tensor products of eigenvectors for each local $\rm{SU(2)}$ operation $U_{k}$ in the respective one-qubit Hilbert spaces.
We adopt the notation $|{+1}\rangle$ and $|{-1}\rangle$ for the local eigenvectors of each $U_{k}$ corresponding to the eigenvalues $e^{i\phi_{k}}$ and $e^{-i\phi_{k}}$, respectively. The eigenvectors of $U$ are then on the form $\bigotimes_{k=1}^{n}|{l_{jk}}\rangle$, where $l_{jk}$ is either $+1$ or $-1$.
In this notation, assuming that the state $|{\psi}\rangle$ is a linear combination of $m$ such eigenvectors, we can write

\begin{eqnarray}\label{exp}|{\psi}\rangle=\sum_{j=1}^{m}b_{j}\bigotimes_{k=1}^{n}|{l_{jk}}\rangle,\nonumber\\
\phantom{yyy}b_{j}\in\mathbb{C}\backslash\{0\},\phantom{yyy}\sum_{j=1}^{m}|{b_{j}}|^2=1.\end{eqnarray}
Any state in $\mathcal{O}^{{\rm{SU(2)}}^{\otimes{n}}}_{|{\psi}\rangle}$ can be expressed on this form in some basis. The following discussion is therefore not dependent on the choice of state in $\mathcal{O}^{{\rm{SU(2)}}^{\otimes{n}}}_{|{\psi}\rangle}$.

Using Eq.~(\ref{exp}), Eq.~(\ref{eq1}) for the ${\rm{SU(2)}}^{\otimes{n}}$-operator $U$ can be split up in $m$ equations for the different eigenvectors. Explicitly, this reads

\begin{eqnarray}\label{bryt}\bigotimes_{k=1}^{n}e^{il_{jk}\phi_{k}}|{l_{jk}}\rangle=e^{i(\chi+2\pi{a_{j}})}\bigotimes_{k=1}^{n}|{l_{jk}}\rangle,\phantom{y}{j=1,\dots,{m}},\nonumber\\\end{eqnarray}
where $e^{i(\chi+2\pi{a_{j}})}$ is the diagonal entry of $U$ belonging to the eigenvector $\bigotimes_{k=1}^{n}|{l_{jk}}\rangle$, and $a_{j}\in{\mathbb{Z}}$ are the winding numbers of the phase factors. The solutions of these equations for $\chi$ gives us all cyclic evolutions whose corresponding stabilizers belong to the same Cartan subgroup. The solutions $\chi$ of Eq.~(\ref{bryt}) depend only on the $a_{j}$'s and which set of eigenvectors that have nonzero coefficients $b_{j}$ but not on the precise values of the $b_{j}$'s. As mentioned before, to find the full set of different phases $\chi$ associated with the ${\rm{SU(2)}}^{\otimes{n}}$-orbit of a state $|{\psi}\rangle$, we must consider all Cartan subgroups, or equivalently solve Eq.~(\ref{bryt}) in every basis. However, since the coefficients $b_{j}$ do not matter for the solutions, we need only consider the different sets of eigenvectors that $|{\psi}\rangle$ can be expanded in. Given such a set, the winding numbers a
 re what tells the different solutions apart.

Equation (\ref{bryt}) defines a system of $m$ linear equations in the exponential coefficients $\sum_{k=1}^{n}l_{jk}\phi_{k}=\chi+2\pi{a_{j}}$. It can  be reexpressed on matrix form as

\begin{eqnarray}\label{yutt}\left(\begin{array}{ccccc}
l_{11} & {l}_{12} & \cdots & {l}_{1n} & -1\\
l_{21} & {l}_{22} & \cdots & {l}_{2n} &-1\\
\vdots & \vdots & \ddots &\vdots & \vdots\\
l_{m1} & {l}_{m2}& \cdots & {l}_{mn}& -1\\
\end{array}\right)\left(\begin{array}{c}
\phi_{1}  \\
\phi_{2}\\
\vdots\\
\phi_{n}\\
\chi\\
\end{array}\right)=\left(\begin{array}{c}
2\pi{a}_{1}\\
2\pi{a}_{2}\\
\vdots\\
2\pi{a}_{m}\\
\end{array}\right).\nonumber\\
\end{eqnarray}
We will refer to the $m\times{(n+1)}$ matrix on the left-hand side of this equation as $A$. This naming is with reference to the $A$-classes previously discussed since the solutions $\chi$ of Eq.~(\ref{yutt}) are solutions for every state in the common $A$-class of all diagonal stabilizers corresponding to the solutions $\phi_{k}$.

Equation (\ref{yutt}) has a solution if it is consistent, i.e., if the $m$ linear equations defined by the rows of $A$ and the corresponding $a_{j}$'s have a common solution. For every $A$-matrix, there are $a_{j}$'s such that the equation is consistent.
If $\chi$ is to take only discrete values it must be uniquely defined for every consistent choice of $a_{j}$. For this to be the case the $A$-matrix must satisfy that the submatrix $\tilde{A}$ made up of the $n$ first columns contains a set $S$ of linearly dependent rows. We can view the rows of $\tilde{A}$ as vectors in $\mathbb{Z}^n$. Linear dependence means that there is a linear combination with coefficients $c_{j}$ of the vectors in $S$ summing to the zero vector. Performing this summation corresponds to a Gauss elimination to put $A$ on a row echelon form. This determines uniquely $\chi$ if $\sum_{j\in{S}}c_{j}\neq0$. The condition under which $\chi$ takes only a discrete set of values for the cyclic evolutions, corresponding to stabilizers in the Cartan subgroup of which $\bigotimes_{k=1}^{n}|{l_{jk}}\rangle$ are eigenvectors, is thus that there exist integers such that

\begin{eqnarray}\label{rewt}\sum_{j\in{S}}c_{j}l_{jk}=0,\phantom{u}\forall{k},\phantom{rr}\sum_{j\in{S}}c_{j}\neq0.\end{eqnarray}
Here, we restrict the $c_{j}$'s to be integers even though linear dependence only requires them to be real nonzero numbers. This is done because any set of $c_{j}$'s, that is a solution of Eq.~(\ref{rewt}), is equivalent up to a common multiplicative factor to a set of integers, and, as we shall see below, this common multiplicative factor is not important to determine the value of $\chi$.

Let us now consider the relation between the vectors $\bigotimes_{k=1}^{n}|{l_{jk}}\rangle$, in Hilbert space, and the corresponding row vectors of $\tilde{A}$.
We choose the set of $n$ generators of the Lie algebra of the Cartan subgroup as the Hermitean operators $\hat{e}_{h}\equiv\hat{1}\otimes\hat{1}\otimes\dots\otimes{s}^{h}\otimes\dots\otimes\hat{1}$, where ${s}^{k}$ is the operation on the $k$th qubit such that ${s}^k|{l_{jk}}\rangle=l_{jk}|{l_{jk}}\rangle$, for  $k=1,\dots,n$. With this choice $\hat{e}_{h}\bigotimes_{k=1}^{n}|{l_{jk}}\rangle=l_{jh}\bigotimes_{k=1}^{n}|{l_{jk}}\rangle$, i.e., the $\{l_{jk}\}$ are the eigenvalues of the $n$ generators $\hat{e}_{h}$ acting on the $j$th basis vector in the expansion of the state $|\psi\rangle$.
The row vectors of $\tilde{A}$ whose elements are $l_{jk}$, thus contains the information about the eigenvalues of the Lie algebra and are called the weight-vectors of $|\psi\rangle$ in the given basis. To get a geometric picture of the set $S$ of weight-vectors we note that they are a subset of the vectors defining the corners of an $n$-dimensional hypercube with side length $2$, centered at the $0$-vector.

The first condition of Eq.~(\ref{rewt}) can be equivalently formulated as the requirement that the subset of row vectors $S$ is such that the $0$-vector is included in the affine hull of $S$. We therefore call states satisfying Eq.~(\ref{rewt}) {\it affinely balanced} states, or $a$-states. It is important to stress that the property of affine balancedness is relative to a Cartan subgroup and thus basis dependent. If only a part of the rows need to be included in $S$ we call the state {\it reducible} in the given basis. Furthermore, if all rows must be included in $S$ we call the state {\it irreducible} in the given basis.

There is a stronger requirement on the state, namely that

\begin{eqnarray}\label{rewt2}\sum_{j\in{S}}c_{j}l_{jk}=0,\phantom{u}\forall{k},\phantom{u}c_{j}>0.\end{eqnarray}
This can equivalently be formulated as that the subset of row vectors $S$ is such that the $0$-vector is included in the convex hull of $S$. We call the states satisfying Eq.~(\ref{rewt2}) {\it convexely balanced} states, or $c$-states, in the given basis. Again, we say that a $c$-state is reducible in the given basis if only a part of the rows needs to be included in $S$, and irreducible in the given basis if all rows must be included.

The states are called balanced with reference
to the observation in Ref. \cite{koff} that $3$-tangle for a three-qubit state is nonzero only if the set of weight vectors, corresponding to corners of a cube, is such that the "center of mass" of the cube is in its convex hull. This terminology was introduced in Ref. \cite{ost} in the context of multipartite entanglement.

So far we have only addressed the question concerning which states have a discrete set of phase factors under the action of a specific Cartan subgroup of ${\rm{SU(2)}}^{\otimes{n}}$. Generally the property that one Cartan subgroup gives a discrete set of phase factors does not imply this for all other Cartan subgroups, which is necessary for the existence of topological phases. In fact, every state is an $a$-state in some basis, and even a $c$-state in some basis. We must therefore address the question of which states can be shown to be $a$-states, or $c$-states in every basis.

To do this we return to the question of ${\rm{SL(2,\mathbb{C})}}^{\otimes{n}}$-semistability.
A state is ${\rm{SL(2,\mathbb{C})}}^{\otimes{n}}$-semistable if and only if the $0$-vector is contained in the convex hull of the weight-vectors of $|\psi\rangle$ for every Cartan subgroup of ${\rm{SL(2,\mathbb{C})}}^{\otimes{n}}$ \cite{dolg}. This trivially implies that the $0$-vector is contained in the convex hull of the weight-vectors of $|\psi\rangle$ for every Cartan subgroup of ${\rm{SU(2)}}^{\otimes{n}}$, and thus an ${\rm{SL(2,\mathbb{C})}}^{\otimes{n}}$-semistable state is a $c$-state in every basis and has topological phases. Note that for all $c$-states, $\chi=\pi$ is a solution of Eq.~(\ref{yutt}). Therefore all ${\rm{SL(2,\mathbb{C})}}^{\otimes{n}}$-semistable states have the topological phase $\pi$ and their ${\rm{SU(2)}}^{\otimes{n}}$-orbits are thus not simply connected.

It was shown by Osterloh and Siewert \cite{ost} that every irreducible $c$-state is ${\rm{SL(2,\mathbb{C})}}^{\otimes{n}}$-equivalent to a state with every reduced one-qubit density matrix maximally mixed, and therefore these states are ${\rm{SL(2,\mathbb{C})}}^{\otimes{n}}$-semistable. Thus, the states that are irreducible $c$-states in some basis have topological phases. For four qubits it was pointed out in Ref.~\cite{ossy} that all irreducible $c$-states belong to the $G_{abcd}$ family of SLOCC-classes.

Finally, we note that the ${\rm{SL(2,\mathbb{C})}}^{\otimes{n}}$-unstable states with topological phases are the states that are $a$-states in every basis but not $c$-states in every basis.

\subsection{Irreducible states of maximal length}

For the purpose of finding the set of topological phases that can result from cyclic evolutions of an $n$-qubit system we can limit our search by considering only a specific class of states. To see this we use linear dependence properties of the rows and columns of the $\tilde{A}$-matrix.

As described above the topological phase $\chi$ is uniquely determined by any set $S$ of rows of $\tilde{A}$ satisfying Eq.~(\ref{rewt}). The smallest possible number of rows of $\tilde{A}$ that satisfies Eq.~(\ref{rewt}) is the minimal number of linearly dependent rows of $\tilde{A}$. Such a smallest possible set $S_{min}$ alone will uniquely determine $\chi$ given a consistent choice of the $\{a_{j}\}$. Any $a$-state has a minimal set of linearly dependent rows, and if it has no rows in addition to this it is irreducible.
For a state of an $n$-qubit system, the minimal number of linearly dependent rows of $\tilde{A}$ is at most $n+1$, and if this is the case it follows that $\rank(A)=n+1$. If, in addition to this, the total number of rows of $\tilde{A}$ is $n+1$ we say that the state is an {\it irreducible $a$-state of maximal length}.

If the $a$-state is such that $A$ has rank $n+1$, whether or not it is irreducible, Eq.~(\ref{yutt}) has a uniquely determined solution not only for $\chi$ but also for $\{\phi_{k}\}$. This means the set of stabilizers $U\in{\rm{SU(2)}}^{\otimes{n}}$, such that $U|\psi\rangle=e^{i\chi}|\psi\rangle$ for each $\chi$, is discrete. If $\rank(A)<n+1$, there is a continuous family of $U$ parametrized by $n+1-\rank(A)$ real parameters for each $\chi$.
If the minimal number of linearly dependent rows is $k$, the submatrix $\tilde{A}_{min}$ of $\tilde{A}$ that consists of these rows will have rank $(k-1)$. Therefore it is possible to remove $n-k+1$ columns from $\tilde{A}_{min}$ and still have the same unique solution for $\chi$ since only $k-1$ columns of $\tilde{A}_{k}$ will be linearly independent. The remaining $k-1$ columns of $\tilde{A}_{min}$ would then correspond to an irreducible $a$-state of maximal length for $(k-1)$ qubits if $k>2$.
If $k=2$, this procedure would produce a single qubit state. If, in this case, we instead remove only $n-k$ columns we would produce an entangled two-qubit state.
Thus, any irreducible $a$-state $|\psi\rangle$ of length $k>2$, in an $n$-qubit system, can be generated from an irreducible $a$-state $|\psi_{(k-1)}\rangle$ of maximal length, in a $(k-1)$-qubit system, by adding $n-k+1$ columns to the $\tilde{A}$-matrix of $|\psi_{(k-1)}\rangle$ from the set of columns with elements $+1$ and $-1$ that are in the linear span of the columns of $\tilde{A}$. This procedure has been termed telescoping \cite{ost}. For two qubits there are no irreducible states of maximal length and hence there are no irreducible states of length three for any number of qubits.
Furthermore, any $a$-state can be generated from an irreducible $a$-state by adding rows to the $\tilde{A}$-matrix in a consistent way. This state will share the set of stabilizers in the Cartan subgroup with the irreducible state.

From the above argument we can see that for the purpose of finding the possible values of $\chi$ in an $n$-qubit system, we need to consider only the irreducible $a$-states of maximal length $k$, for $k=4,\dots,n+1$ and in addition to this the irreducible states of two qubits.

Finding the $\chi$ for an irreducible $n$-qubit $a$-state of maximal length associated with evolutions that belong to the Cartan subgroup of ${\rm{SU(2)}}^{\otimes{n}}$ corresponds to reducing $A$ to row echelon form by adding up rows multiplied by the integers ${c}_{0},{c}_{1},{c}_{2},\dots,{c}_{n}$, henceforth indexed such that $|{c}_{0}|\geq|{c}_{1}|\geq|{c}_{2}|\geq\dots\geq|{c}_{n}|$. Given these integers the equation for $\chi$ is

\begin{eqnarray}-\chi\sum_{j=0}^{n}{c_{j}}=2\pi\sum_{j=0}^{n}{c_{j}a_{j}}\Rightarrow\chi=-\frac{2\pi\sum_{j=0}^{n}{c_{j}a_{j}}}{\sum_{j=0}^{n}{c_{j}}}.\end{eqnarray}
Since the coefficients $c_{j}$ are uniquely defined by $A$ up to a common factor, there is a unique choice of $c_{j}$ as integers without common prime factor, and the solution of $\chi$ is a rational number times $2\pi$. Given that the $c_{j}$ are chosen without common prime factor, the minimal nonzero topological phase $\chi_{min}$ for a given irreducible $a$-state, and free choice of the parameters $a_{j}$, is

\begin{eqnarray}\chi_{min}=\frac{2\pi}{\sum_{j=0}^{n}{c_{j}}}.\end{eqnarray}
Note that $\sum_{j=0}^{n}{c_{j}}$ is always an even number.

In Sec. \ref{sec52} we saw that all irreducible $c$-states have topological phases. Furthermore, we note that every $a$-state can be related to a $c$-state by changing the signs of the rows in $\tilde{A}$ that correspond to negative coefficients $c_{j}$. This $c$-state has by definition all $c_{j}$ positive and therefore the sum $\sum_{j=0}^{n}{c_{j}}$ will always be greater than for the corresponding $a$-state. Thus, the set of $n$-qubit $c$-states contains the states with smallest possible $\chi_{min}$ for an $n$-qubit system.

From the present analysis we see that if, for each irreducible $c$-state of maximal length in a $k$-qubit system, and, for each $k=3,\dots,n$, we calculate the topological phases for the Cartan subgroup of ${\rm{SU(2)}}^{\otimes{k}}$, of diagonal operators in the basis in which the state is expressed, we will get an exhaustive list of possible topological phases in an $n$-qubit system.

\subsection{Combinatorial reformulation}

\label{redc}The problem of finding the topological phases for the irreducible states of maximum length can be reformulated as a combinatorial problem. To see this, let us denote the multiset of coefficients $c_{j}$ by $P$.
In reducing the $A$-matrix to row echelon form the elements of each of the  columns of $\tilde{A}$ are multiplied by the coefficients $c_{j}$ and summed to zero. For each column $k$, a given coefficient will be assigned to either a multiset $P^{+}_{k}$ if it multiplies $+1$, or to a multiset $P^{-}_{k}$ if it multiplies $-1$. Therefore for each column of $\tilde{A}$, $P^{+}_{k}$ and $P^{-}_{k}$ defines an equal sum partition of $P$.

To proceed we note that different states for which the respective $A$-matrices differ only by a multiplication of the $k$th column of $A$ by $-1$, for $1\leq{k}\leq{n}$, have the same solutions for $\chi$. Furthermore, for any state, an operation on the $k$th qubit that interchanges $|+1\rangle$ and $|-1\rangle$ corresponds to multiplication by $-1$ in the $k$th column of the corresponding $A$ matrix. Therefore multiplication of any of the first $n$ columns of $A$ corresponds to a change of basis, or alternatively an ${\rm{SU(2)}}^{\otimes{n}}$-operation. Permuting rows of $A$ does not change the solutions and has no physical meaning at all. For calculational purposes we can therefore select the $A$-matrix where the first row corresponds to the largest positive coefficient $c_{0}$ and let the first $n$ elements of this row be $+1$. The other rows are ordered such that the $j$th row correspond to $c_{j-1}$.
By this convention above we have assigned $c_{0}$ to $P^{+}_{k}$ for every $k$.

The coefficients $c_{j}$ are uniquely defined by their memberships in the multisets $P^{+}_{k}$ and $P^{-}_{k}$. In fact, the $c_{j}$'s are uniquely defined by the multisets $P^{+}_{k}\backslash\{c_{0}\}$ alone. This can be understood by removing the first row of $\tilde{A}$ and consider the columns of the remaining matrix as the rows of an equation

\begin{eqnarray}\label{col}\left(\begin{array}{cccc}
l_{21} & {l}_{31} & \cdots & {l}_{(n+1)1}  \\
l_{22} & {l}_{32} & \cdots & {l}_{(n+1)2} \\
\vdots & \vdots & \ddots &\vdots \\
l_{2n} & {l}_{3n}& \cdots & {l}_{(n+1)n}\\
\end{array}\right)\left(\begin{array}{c}
c_{1}  \\
c_{2}\\
\vdots\\
c_{n}\\
\end{array}\right)=-c_{0}\left(\begin{array}{c}
1  \\
1\\
\vdots\\
1\\
\end{array}\right).
\end{eqnarray}
Here, the rows of the left-hand side uniquely determines the coefficients $c_{j}$ up to a common multiplicative factor. Each row, containing only $+1$'s and $-1$'s, is in turn uniquely determined by a specification of which elements are $+1$, i.e., $P^{+}_{k}\backslash\{c_{0}\}$.

Therefore, the problem of finding the topological phases for $n$-qubit irreducible $a$-states of maximal length can be reformulated as the problem of finding multisets of integers with greatest common divisor 1, containing $n$ elements, such that there exist at least $n$ equal sum proper submultisets and such that the integers are uniquely defined by their memberships in these submultisets.

We can make a few useful observations about the multisets $\{c_{1},\dots,c_{n}\}$.
Let $r$ be the number of distinct integers in $\{c_{1},\dots,c_{n}\}$. If we consider the case where $r=n$, i.e., all elements of $\{c_{1},\dots,c_{n}\}$ are distinct, we can see that for $n\leq6$ there is no such solutions. This follows from the fact that no set of positive real integers can have a greater number of equal sum subsets than does $\{1,2,3,\dots,{n}\}$ \cite{stan}. The maximal number of equal sum subsets for $n=1,2,3,\dots$ are $1,1,2,2,3,5,8,14,23,40,\dots$. Thus for $n\geq{7}$ we need to consider this possibility.

Several multisets $\{c_{1},\dots,c_{n}\}$ correspond to the same multiset $P$, and therefore to the same states. Requiring that $|c_{0}|\geq|c_{1}|$ imposes a restriction that avoids this redundancy.

An $A$-matrix for $n$-qubits with nontrivial phases, where the corresponding $\{c_{1},\dots,c_{n}\}$ has $r$ distinct elements, can always be used to generate an $A$-matrix for $(n+1)$ qubits with $r$ distinct elements by increasing the multiplicity of one of the elements of $\{c_{1},\dots,c_{n}\}$. On the other hand $A$-matrices for $(n+1)$ qubits can in general not be used to generate $A$-matrices for $n$ qubits by decreasing multiplicities. Therefore the number of different $A$-matrices is non-decreasing with increasing $n$.

\section{Computational results}

For a \label{ytrrium}small number of qubits the method outlined in Sec. \ref{sec5} can be used to quickly find the full set of topological phases. However, the number of states, or equivalently the number of combinatorial structures, that need to be tested grows rapidly with $n$.
Therefore, a search algorithm, for arbitrary $n$, based on the combinatorial formulation of the problem, has been created. The algorithm has been used to find all combinatorial structures corresponding to $A$-matrices of irreducible $c$-states of maximal length, for up to seven qubits.
For each structure the minimal topological phase $\chi_{min}$ has been calculated. The result is presented in Table \ref{table}. The time required to complete the search increases rapidly with $n$.
Therefore, for eight qubits only partial results have been obtained.
These show that the smallest $\chi_{min}$ is equal to or smaller than $\frac{\pi}{38}$.

\begin{table*}[tb]
\begin{tabular*}{14cm}{r p{2cm} l}
  \hline\hline
  \phantom{i}$n$\phantom{i} & & The set of $\chi_{min}$ of an $n$-qubit system \\[1.0ex]
\hline
  \phantom{i}$2$\phantom{i} & &  $\pi$ \\[1.0ex]
  \phantom{i}$3$\phantom{i} & &  \large{\small{$\pi$}} $\frac{\pi}{2}$\\[1.0ex]
  \phantom{i}$4$\phantom{i} & &  \large{\small{$\pi$}} $\frac{\pi}{2}$ $\frac{\pi}{3}$\\[1.0ex]
\phantom{i}$5$\phantom{i} & &  \large{\small{$\pi$}} $\frac{\pi}{2}$ $\frac{\pi}{3}$ $\frac{\pi}{4}$ $\frac{\pi}{5}$ \\[1.0ex]
\phantom{i}$6$\phantom{i} & &  \large{\small{$\pi$}} $\frac{\pi}{2}$ $\frac{\pi}{3}$ $\frac{\pi}{4}$ $\frac{\pi}{5}$ $\frac{\pi}{6}$ $\frac{\pi}{7}$ $\frac{\pi}{8}$ $\frac{\pi}{9}$\\[1.0ex]
\phantom{i}$7$\phantom{i} & &  \large{\small{$\pi$}} $\frac{\pi}{2}$ $\frac{\pi}{3}$ $\frac{\pi}{4}$ $\frac{\pi}{5}$ $\frac{\pi}{6}$ $\frac{\pi}{7}$ $\frac{\pi}{8}$ $\frac{\pi}{9}$ $\frac{\pi}{10}$ $\frac{\pi}{11}$ $\frac{\pi}{12}$ $\frac{\pi}{13}$ $\frac{\pi}{14}$ $\frac{\pi}{15}$ $\frac{\pi}{16}$ $\frac{\pi}{17}$ $\frac{\pi}{18}$\\[1.0ex]

  \hline\hline
\end{tabular*}
\caption{Results of the search algorithm. The set of minimal topological phases $\chi_{min}$ corresponding to the set of irreducible $c$-states of $n$ qubits for $n=2,3,4,5,6,7$. Integer multiples of these $\chi_{min}$ are the only topological phases for respective number of qubits $n$.}
\label{table}
\end{table*}

As expected from the discussion in Sec. \ref{ent} the only $\chi_{min}$ for two and three qubits are $\pi$ and $\{\pi,\frac{\pi}{2}\}$, respectively.
For four qubits it was found that the set of $\chi_{min}$ is $\{\pi,\frac{\pi}{2},\frac{\pi}{3}\}$.
In Sec. \ref{ent}, we saw that the existence of a topological phase $\frac{2\pi}{d}$ implies the existence of a generator of the algebra of ${\rm{SL(2,\mathbb{C})}}^{\otimes{n}}$-invariant polynomials of degree $md$ for some integer $m$. For two, three and four qubits, each generator of the polynomial invariant algebra of degree $d$ is nonzero on some ${\rm{SU(2)}}^{\otimes{n}}$-orbit, and indeed $A$-class, associated with a topological phase $\frac{2\pi}{d}$.
This is the smallest possible topological phase a state can have, given the degree of the generator.
Thus, for up to four qubits there is a correspondence between the topological phases and the degree of the generating polynomials.
For higher number of qubits, the generators of the algebra of ${\rm{SL(2,\mathbb{C})}}^{\otimes{n}}$-invariant polynomials are not fully known.
However, it was conjectured by Luque and Thibon \cite{luk} that the degree of the generators of the invariant algebra for five qubits are $4,6,8,10,$ and $12$. Subsequently it was shown by {\DJ}okovi\'c and Osterloh \cite{dok} that generators of degree up to $12$ do indeed exist. This implies that there is no apparent correspondence for arbitrary number of qubits since there is no $\frac{\pi}{6}$ phase, corresponding to a generator of degree $12$, in the case of five qubits. We may still note, for at least up to five qubits, that every state with $\chi_{min}=\frac{2\pi}{d}$ such that it is not a multiple of any other minimal topological phase, is in a SLOCC-class for which a generator of degree $d$ is nonzero.

The lowest number of qubits for which the topological phases has not already been found through the examples of Sec. \ref{sec3}, is five. We give the combinatorial structures and examples of the corresponding states, for this case, as examples of the results found using the combinatorial method.

For five qubits there are only four combinatorial structures satisfying the conditions defined in Sec. \ref{redc}, that are associated with $c$-states. The first one, the multiset $\{1,1,1,1,1\}$ with equal sum submultisets $\{1,1\}$, $\{1,1\}$, $\{1,1\}$, $\{1,1\}$, and $\{1,1\}$ is associated with states for which
$\chi_{min}=\frac{\pi}{3}$. An example of a state associated with this structure, is

\begin{eqnarray}\label{s3}|\psi\rangle&=&|11111\rangle+|10001\rangle+|11000\rangle\nonumber\\
&&+|01100\rangle+|00110\rangle+|00011\rangle.\end{eqnarray}
The next structure $\{1,1,1,1,1\}$ with equal sum submultisets $\{1\}$, $\{1\}$, $\{1\}$, $\{1\}$, and $\{1\}$ corresponds to the states of the class in Eq.~(\ref{yhn}) and has $\chi_{min}=\frac{\pi}{4}$

\begin{eqnarray}\label{s4}|\psi\rangle&=&|11111\rangle+|10000\rangle+|01000\rangle\nonumber\\
&&+|00100\rangle+|00010\rangle+|00001\rangle\nonumber\\
&=&|11111\rangle+|W^{5}\rangle.\end{eqnarray}
The third example is an irreducible five-qubit $c$-state of maximal length with $\chi_{min}=\frac{\pi}{4}$
\begin{eqnarray}\label{s5}|\psi\rangle=&|11111\rangle+|10000\rangle+|01001\rangle\nonumber\\
&+|01100\rangle+|00110\rangle+|00011\rangle,
\end{eqnarray}
that is associated with the combinatorial structure consisting of the multiset $\{2,1,1,1,1\}$ and the equal sum submultisets $\{2\}$, $\{1,1\}$, $\{1,1\}$, $\{1,1\}$, and $\{1,1\}$.
Finally, we consider an irreducible five-qubit $c$-state of maximal length with $\chi_{min}=\frac{\pi}{5}$
\begin{eqnarray}\label{st5}|\psi\rangle&=&|11111\rangle+|10000\rangle+|01000\rangle\nonumber\\
&&+|00110\rangle+|00101\rangle+|00011\rangle,
\end{eqnarray}
and the corresponding combinatorial structure is the multiset $\{2,2,1,1,1\}$ with equal sum submultisets $\{2\}$, $\{2\}$, $\{1,1\}$, $\{1,1\}$, and $\{1,1\}$.

The five-qubit states corresponding to the other topological phases in Table \ref{table}, i.e., the irreducible $c$-states with $\chi_{min}=\pi,\frac{\pi}{2}$ and $\frac{\pi}{3}$ can be constructed from from irreducible states of fewer qubits by duplicating columns of the $\tilde{A}$-matrix.

A complete list of the multisets found by the algorithm, including the case of six and seven qubits, and the associated topological phases is presented in \cite{supp}. Here we give one more example to illustrate a special case.
As mentioned in Sec. \ref{redc} it is only for seven or more qubits that there are combinatorial structures with multisets where all elements are distinct.
An example of such a seven-qubit state is

\begin{eqnarray}\label{st18}|\psi\rangle&=&|1111111\rangle+|1100000\rangle+|0011000\rangle\nonumber\\
&&+|0000110\rangle+|0010101\rangle+|1001011\rangle\nonumber\\&&+|0100011\rangle+|0101101\rangle,
\end{eqnarray}
for which $\chi_{min}=\frac{\pi}{18}$.
The combinatorial structure corresponding to this state is the multiset $\{7,6,5,4,3,2,1\}$ and the seven equal sum  submultisets $\{7,3\}$, $\{7,2,1\}$, $\{6,4\}$, $\{6,3,1\}$, $\{5,4,1\}$, $\{5,3,2\}$, and $\{4,3,2,1\}$.

The states and the associated topological phases found by the algorithm show us which topological phase factors exist in an $n$-qubit system. However, we have not yet touched upon the question of unitary, or SLOCC, equivalence between the states found. From knowledge of the topological phases associated with states we can in many cases determine that the states are unitarily inequivalent. Let us assume that two $n$-qubit states $|\psi_{1}\rangle$ and $|\psi_{2}\rangle$ have topological phases equal to $\frac{\pi}{d_{1}}$ and $\frac{\pi}{d_{2}}$, respectively, and are unitarily equivalent. Then we will be able to combine  $p$ cyclic evolutions corresponding to $\frac{\pi}{d_{1}}$ and $r$ cyclic evolutions corresponding to $\frac{\pi}{d_{2}}$, where $p$ and $r$ are integers, to form a new cyclic evolution. By B\'ezout's identity we can always choose $p$ and $r$ such that

\begin{eqnarray}\frac{p\pi}{d_{1}}+\frac{r\pi}{d_{2}}=\frac{gcd(d_{2},d_{1})\pi}{d_{1}d_{2}}\end{eqnarray}
and this is the smallest possible topological phase that can result from such a combination. Thus if the two states are unitarily equivalent, this topological phase must be a multiple of one of the $\chi_{min}$ associated with the $n$-qubit system. If this is not true the two states are unitarily inequivalent. From before we remember that each irreducible $c$-state is ${\rm{SL(2,\mathbb{C})}}^{\otimes{n}}$-equivalent to a maximally entangled state \cite{ost}, and that there can be at most one maximally entangled state, up to local unitary transformations, in each SLOCC-orbit. Therefore unitary inequivalence of the irreducible $c$-states implies that they are also SLOCC inequivalent.

By this argument, we can for example conclude that the five-qubit states corresponding to the phase $\frac{\pi}{3}$ are in different SLOCC-orbits than those corresponding to $\frac{\pi}{4}$ or $\frac{\pi}{5}$. Likewise for six qubits there are different SLOCC-orbits corresponding to the topological phases $\frac{\pi}{5},\frac{\pi}{6},\frac{\pi}{7},\frac{\pi}{8}$, and $\frac{\pi}{9}$ respectively. For seven qubits, the states corresponding to the topological phases $\frac{\pi}{10},\frac{\pi}{11},\frac{\pi}{12},\frac{\pi}{13},\frac{\pi}{14},\frac{\pi}{15},\frac{\pi}{16},\frac{\pi}{17}$, and $\frac{\pi}{18}$ are all mutually inequivalent under SLOCC.

\section{Conclusions}

Topological phases originate from the non-trivial topological structure of the ${\rm{SU(2)}}^{\otimes{n}}$-orbits of entangled multi-qubit states and are thus an attribute of multipartite entanglement. The characterization in terms of equivalence classes of states with the same topological phase, is intermediate in coarseness between SLOCC-equivalence and local unitary equivalence. We have discussed the relationship between topological phases and entanglement properties, in particular the relation to nonzero SLOCC-invariant polynomials.
It was found that for up to four qubits there is a direct correspondence between the topological phase factors and the degree of the generating polynomials of the algebra of polynomial SLOCC-invariants.
We have shown that all ${\rm{SL(2,\mathbb{C})}}^{\otimes{n}}$-semistable $n$-qubit states have topological phases and thus that their ${\rm{SU(2)}}^{\otimes{n}}$-orbits are not simply connected. We have  demonstrated that there are states with topological phases in the class of ${\rm{SL(2,\mathbb{C})}}^{\otimes{n}}$-unstable states for any number of qubits.

We have described a method for finding the topological phases for an arbitrary number of qubits and used this method to find the full set of topological phases for up to seven qubits. While the ${\rm{SU(2)}}^{\otimes{2}}$-orbit of an entangled qubit pair is doubly connected, for higher number of qubits the presence of phases that are smaller fractions of $\pi$ implies that the corresponding orbits are multiply connected, i.e, have highly non-trivial topology.
The results show that the number of different topological phases grows rapidly with the number of qubits.

Several open questions remain.

(1) The relationship between topological phases and the degree of the generating polynomials is unclear. A particular question is if it is true, for an $n$-qubit system, that every state with $\chi_{min}=\frac{2\pi}{d}$ such that it is not a multiple of any other minimal topological phase, is in a SLOCC-class for which a generator of degree $d$ is nonzero. If true, these topological phases could be used to predict the degree of some of the generating polynomials.

(2) The question of the relation between ${\rm{SU(2)}}^{\otimes{n}}$-invariant polynomials and topological phases has only been touched upon. In particular, the question regarding the correspondence between topological phases and the degree of the generating polynomials is pertinent.

(3) In the case of two, three, and four qubits a randomly selected state has unit probability to have only the topological phase $\pi$. Thus, states with topological phases other than $\pi$ is a set of zero measure. It is unknown whether this is a general behavior for larger number of qubits.

(4) No general criteria regarding existence of topological phases in ${\rm{SL(2,\mathbb{C})}}^{\otimes{n}}$-unstable states have been found. The existence of nonzero ${\rm{SU(2)}}^{\otimes{n}}$-invariant polynomials for this category of states would imply the existence of topological phases. The question is thus if such invariants exist and if their existence is related to a criterion similar to that of semistability.

Finally, a natural extension of this work would be to consider cyclic evolutions in the full ${\rm{SL(2,\mathbb{C})}}^{\otimes{n}}$ group. Besides allowing $\chi$ to be a complex phase, it would extend the set of cyclic evolutions to those corresponding to non-diagonalizable stabilizers, which cannot be analyzed by the method used in this paper. These cyclic evolutions would give a set of topological phases related to the topological structure of the given ${\rm{SL(2,\mathbb{C})}}^{\otimes{n}}$-orbit. While all ${\rm{SL(2,\mathbb{C})}}^{\otimes{n}}$-semistable orbits have such topological phases, it is unclear to us if there exist ${\rm{SL(2,\mathbb{C})}}^{\otimes{n}}$-unstable orbits with topological phases.

For the ${\rm{SL(2,\mathbb{C})}}^{\otimes{n}}$-semistable states only real-valued $\chi$ can exist, since no polynomial in the expansion coefficients of the state is invariant under multiplication by a complex number with absolute value different from unity. This is directly related to the fact that all cyclic ${\rm{SL(2,\mathbb{C})}}^{\otimes{n}}$-evolutions are norm-preserving for this class of states.
For less than five qubits these norm-preserving cyclic evolutions  would not yield any additional topological phases to those given in Table \ref{table}, since the degree of the generating ${\rm{SL(2,\mathbb{C})}}^{\otimes{n}}$-invariant polynomials forbids it. For five or more qubits additional phases for ${\rm{SL(2,\mathbb{C})}}^{\otimes{n}}$ evolutions could be a possibility.

\section*{Acknowledgments}
We thank Andreas Osterloh for pointing out Refs.~\cite{dok} and \cite{ossy} as well as for useful correspondence.
M.E. acknowledges support from the Swedish Research Council (VR).
E.S. and M.S.W. acknowledge support from the National Research Foundation and the Ministry of
Education (Singapore).

\section{Supplemental material}

The problem of finding topological phases for an $n$-qubit system has been reduced to the problem of finding multisets of integers $\{c_{1},\dots,c_{n}\}$ with greatest common divisor 1 such that there exist at least $n$ equal sum proper submultisets, and such that integers are uniquely defined by their memberships in these submultisets. In addition to this the multisets are required to satisfy that $|c_{0}|\geq|c_{1}|$.
Here we present a complete list of all such multisets of positive integers that correspond to irreducible $c$-states of maximal length for $n=3,\dots,7$. The two-qubit is not included since there are no irreducible $c$-states of maximal length for two-qubits.

From these multisets the corresponding topological phases and $A$-classes can be constructed.
Given a multiset $\{c_{1},\dots,c_{n}\}$ where the $n$ equal sum submultisets each have the sum $Z$, the minimal topological phase $\chi_{min}$ is uniquely determined by the sum $\sum_{j=1}^{n}c_{j}$ and $Z$ as $\chi_{min}=\frac{\pi}{(\sum_{j=1}^{n}c_{j}-Z)}$. As we are only interested in $c$-states where we require the $c_{j}$'s are all positive, we must add the condition that $Z<\frac{1}{2}\sum_{j=1}^{n}c_{j}$, since otherwise $c_{0}$ is negative.

For many multisets there is more then one way to choose $n$ equal sum submultisets with a given sum $Z$ such that the integers are still uniquely defined by their memberships in these submultisets. Each way of choosing the $n$ submultisets correspond to an $A$-class of states up to permutations of the qubits. Therefore, for a given multiset and a given $Z$ there can be many $A$-classes.
Because of this, we do not present a list of all $A$-classes but instead list only the multisets and $Z$ together with the corresponding $\chi_{min}$.

To find the $A$-classes corresponding to a specific multiset $\{c_{1},\dots,c_{n}\}$ and $Z$, one can identify the different choices of submultisets. These submultisets can then be used to construct the rows of the following equation

\begin{eqnarray}\label{col}\left(\begin{array}{cccc}
l_{21} & {l}_{31} & \cdots & {l}_{(n+1)1}  \\
l_{22} & {l}_{32} & \cdots & {l}_{(n+1)2} \\
\vdots & \vdots & \ddots &\vdots \\
l_{2n} & {l}_{3n}& \cdots & {l}_{(n+1)n}\\
\end{array}\right)\left(\begin{array}{c}
c_{1}  \\
c_{2}\\
\vdots\\
c_{n}\\
\end{array}\right)=-c_{0}\left(\begin{array}{c}
1  \\
1\\
\vdots\\
1\\
\end{array}\right).
\end{eqnarray}
If the resulting matrix on the right-hand side has nonzero determinant, it corresponds to an irreducible $c$-state of maximal length.

As an example we can consider the multiset $\{4,3,3,1,1,1,1\}$ for which there is ten proper submultisets with sum $4$. These are $\{4\}$, $\{3,1\}$, $\{3,1\}$, $\{3,1\}$, $\{3,1\}$, $\{3,1\}$, $\{3,1\}$, $\{3,1\}$, $\{3,1\}$, and $\{1,1,1,1\}$. From these submultisets we need to choose seven corresponding to a choice of rows in Eq. (\ref{col}). However, not any selection of seven out of these submultisets will be such that the integers are uniquely defined by their memberships in the submultisets. To see this let us temporarily obscure the identities of the integers and denote them as $\{A,B,B,C,C,C,C\}$ and the submultisets as $\{A\}$, $\{B,C\}$, $\{B,C\}$, $\{B,C\}$, $\{B,C\}$, $\{B,C\}$, $\{B,C\}$, $\{B,C\}$, $\{B,C\}$, and $\{C,C,C,C\}$
We then see that $\{A\}$ and $\{C,C,C,C\}$ need to be included in the selection since $A=4$ is the only equation that defines the integer $A$, and the equation $B+C=4$ does not uniquely define $B$ or $C$ unless in  conjunction with $C+C+C+C=4$.

Given that we have found a selection of submultisets that uniquely defines the integers we can create the corresponding matrix. For example, a choice corresponding to $\{4\}$, $\{3,1\}$, $\{3,1\}$, $\{3,1\}$, $\{3,1\}$, $\{3,1\}$, and $\{1,1,1,1\}$ is

\begin{eqnarray}
\left(\begin{array}{ccccccc}
+1 & -1 & -1 & -1 & -1 & -1 & -1  \\
-1 & +1 & -1 & +1 & -1 & -1 & -1  \\
-1 & +1 & -1 & -1 & +1 & -1 & -1   \\
-1 & +1 & -1 & -1 & -1 & +1 & -1   \\
-1 & +1 & -1 & -1 & -1 & -1 & +1   \\
-1 & -1 & +1 & -1 & -1 & +1 & -1   \\
-1 & -1 & -1 & +1 & +1 & +1 & +1   \\
\end{array}\right)\left(\begin{array}{c}
4  \\
3\\
3\\
1\\
1\\
1\\
1\\

\end{array}\right)=-6\left(\begin{array}{c}
1  \\
1\\
1\\
1\\
1\\
1\\
1\\
\end{array}\right).\nonumber\\
\end{eqnarray}
Identifying the columns of the matrix on the left-hand side, and the column vector on the right-hand side with the weight-vectors of the state we can easily write down a representative of the $A$-class as

\begin{eqnarray}|\psi\rangle&=&|1111111\rangle+|1000000\rangle+|0111100\rangle\nonumber\\
&&+|0000010\rangle+|0100001\rangle+|0010001\rangle\nonumber\\&&+|0001011\rangle+|0000101\rangle.
\end{eqnarray}

The multisets for three, four, five, and six qubits are given in Table. \ref{table1}, Table. \ref{table2}, Table. \ref{table3}, and Table. \ref{table4} respectively. The multisets corresponding to seven qubits have been divided into Tables \ref{table5} to \ref{table10} based on how many distinct integers they contain for readability.

\begin{table}[!b]
\begin{tabular*}{8.6cm}{c c c}
  \hline\hline
  \phantom{i}multiset\phantom{i} & \phantom{i}sum of $P_{+}\backslash{c_{0}}$\phantom{i}& \phantom{i}$\chi_{min}$\phantom{i} \\[1.0ex]
\hline
\phantom{i}$\{1,1,1\}$\phantom{i} & 1 &  \large$\frac{\pi}{2}$ \\[1.0ex]

  \hline\hline

\end{tabular*}
\caption{List of multisets, sums of the equal sum submultisets, and $\chi_{min}$ associated to three-qubit irreducible $c$-states of maximal length.}
\label{table1}
\end{table}

\begin{table}[!tb]
\begin{tabular*}{8.6cm}{c c c}
  \hline\hline
  \phantom{i}multiset\phantom{i} & \phantom{i}sum of $P_{+}\backslash{c_{0}}$\phantom{i}& \phantom{i}$\chi_{min}$\phantom{i} \\[1.0ex]
\hline
\phantom{i}$\{1,1,1,1\}$\phantom{i} & 1 &  \large$\frac{\pi}{3}$ \\[1.0ex]

  \hline\hline

\end{tabular*}
\caption{List of multisets, sums of the equal sum submultisets, and $\chi_{min}$ associated to four-qubit irreducible $c$-states of maximal length.}
\label{table2}
\end{table}

\begin{table}[!tb]

\begin{tabular*}{8.6cm}{c c c}
  \hline\hline
  \phantom{i}multiset\phantom{i} & \phantom{i}sum of $P_{+}\backslash{c_{0}}$\phantom{i}& \phantom{i}$\chi_{min}$\phantom{i} \\[1.0ex]
\hline
\phantom{i}$\{1,1,1,1,1\}$\phantom{i} & 1 &  \large$\frac{\pi}{4}$ \\[1.0ex]
\phantom{i}$\{1,1,1,1,1\}$\phantom{i} & 2 &  \large$\frac{\pi}{3}$ \\[1.0ex]
  \phantom{i}$\{2,1,1,1,1\}$\phantom{i} & 2 &  \large$\frac{\pi}{4}$ \\[1.0ex]
  \phantom{i}$\{2,2,1,1,1\}$\phantom{i} & 2 &  \large$\frac{\pi}{5}$\\[1.0ex]

  \hline\hline

\end{tabular*}
\caption{List of multisets, sums of the equal sum submultisets, and $\chi_{min}$ associated to five-qubit irreducible $c$-states of maximal length.}
\label{table3}
\end{table}

\begin{table}[!tb]

\begin{tabular*}{8.6cm}{c c c}
  \hline\hline
  \phantom{i}multiset\phantom{i} & \phantom{i}sum of $P_{+}\backslash{c_{0}}$\phantom{i}& \phantom{i}$\chi_{min}$\phantom{i} \\[1.0ex]
\hline
\phantom{i}$\{1,1,1,1,1,1\}$\phantom{i} & 1 &  \large$\frac{\pi}{5}$ \\[1.0ex]
\phantom{i}$\{1,1,1,1,1,1\}$\phantom{i} & 2 &  \large$\frac{\pi}{4}$ \\[1.0ex]
 \phantom{i}$\{2,1,1,1,1,1\}$\phantom{i} & 2 &  \large $\frac{\pi}{5}$ \\[1.0ex]
\phantom{i}$\{2,2,1,1,1,1\}$\phantom{i} & 2 &  \large$\frac{\pi}{6}$\\[1.0ex]

  \phantom{i}$\{2,2,2,1,1,1\}$\phantom{i} & 2 &  \large$\frac{\pi}{7}$ \\[1.0ex]
  \phantom{i}$\{2,2,2,1,1,1\}$\phantom{i} & 3 &  \large$\frac{\pi}{6}$ \\[1.0ex]
 \phantom{i}$\{2,2,2,2,1,1\}$\phantom{i} & 4 &  \large$\frac{\pi}{6}$ \\[1.0ex]
\phantom{i}$\{3,3,1,1,1,1\}$\phantom{i} & 3 &  \large $\frac{\pi}{7}$ \\[1.0ex]
\phantom{i}$\{3,2,1,1,1,1\}$\phantom{i} & 3 &  \large$\frac{\pi}{6}$ \\[1.0ex]

\phantom{i}$\{3,2,2,1,1,1\}$\phantom{i} & 3 &  \large$\frac{\pi}{7}$ \\[1.0ex]
\phantom{i}$\{3,2,2,2,1,1\}$\phantom{i} & 4 &  \large$\frac{\pi}{7}$ \\[1.0ex]
\phantom{i}$\{3,3,2,1,1,1\}$\phantom{i} & 3 &  \large$\frac{\pi}{8}$ \\[1.0ex]
\phantom{i}$\{3,3,2,2,1,1\}$\phantom{i} & 4 &  \large$\frac{\pi}{8}$ \\[1.0ex]
\phantom{i}$\{4,2,2,2,1,1\}$\phantom{i} & 4 &  \large$\frac{\pi}{8}$ \\[1.0ex]

\phantom{i}$\{4,3,2,2,1,1\}$\phantom{i} & 4 &  \large$\frac{\pi}{9}$ \\[1.0ex]

  \hline\hline

\end{tabular*}
\caption{List of multisets, sums of the equal sum submultisets, and $\chi_{min}$ associated to six-qubit irreducible $c$-states of maximal length.}
\label{table4}
\end{table}

\begin{table}[!tb]

\begin{tabular*}{8.6cm}{c c c}
  \hline\hline
  \phantom{i}multiset\phantom{i} & \phantom{i}sum of $P_{+}\backslash{c_{0}}$\phantom{i}& \phantom{i}$\chi_{min}$\phantom{i} \\[1.0ex]
\hline
\phantom{i}$\{1,1,1,1,1,1,1\}$\phantom{i} & 1 &  \large$\frac{\pi}{6}$ \\[1.0ex]
\phantom{i}$\{1,1,1,1,1,1,1\}$\phantom{i} & 2 &  \large$\frac{\pi}{5}$ \\[1.0ex]
\phantom{i}$\{1,1,1,1,1,1,1\}$\phantom{i} & 3 &  \large$\frac{\pi}{4}$ \\[1.0ex]

\phantom{i}$\{2,1,1,1,1,1,1\}$\phantom{i} & 2 &  \large $\frac{\pi}{6}$ \\[1.0ex]

  \phantom{i}$\{2,2,1,1,1,1,1\}$\phantom{i} & 2 &  \large $\frac{\pi}{7}$ \\[1.0ex]
\phantom{i}$\{2,2,1,1,1,1,1\}$\phantom{i} & 3 &  \large$\frac{\pi}{6}$ \\[1.0ex]

  \phantom{i}$\{2,2,2,1,1,1,1\}$\phantom{i} & 2 &  \large$\frac{\pi}{8}$\\[1.0ex]
\phantom{i}$\{2,2,2,1,1,1,1\}$\phantom{i} & 3 &  \large$\frac{\pi}{7}$ \\[1.0ex]
\phantom{i}$\{2,2,2,1,1,1,1\}$\phantom{i} & 4 &  \large$\frac{\pi}{6}$ \\[1.0ex]

\phantom{i}$\{2,2,2,2,1,1,1\}$\phantom{i} & 2 &  \large$\frac{\pi}{9}$ \\[1.0ex]
\phantom{i}$\{2,2,2,2,1,1,1\}$\phantom{i} & 3 &  \large$\frac{\pi}{8}$ \\[1.0ex]
\phantom{i}$\{2,2,2,2,1,1,1\}$\phantom{i} & 4 &  \large$\frac{\pi}{7}$ \\[1.0ex]
\phantom{i}$\{2,2,2,2,2,1,1\}$\phantom{i} & 4 &  \large$\frac{\pi}{8}$ \\[1.0ex]

\phantom{i}$\{3,1,1,1,1,1,1\}$\phantom{i} & 3 &  \large$\frac{\pi}{6}$ \\[1.0ex]

\phantom{i}$\{3,3,1,1,1,1,1\}$\phantom{i} & 3 &  \large$\frac{\pi}{8}$ \\[1.0ex]
\phantom{i}$\{3,3,1,1,1,1,1\}$\phantom{i} & 4 &  \large$\frac{\pi}{7}$ \\[1.0ex]

\phantom{i}$\{3,3,3,1,1,1,1\}$\phantom{i} & 3 &  \large$\frac{\pi}{10}$ \\[1.0ex]
\phantom{i}$\{3,3,3,1,1,1,1\}$\phantom{i} & 4 &  \large$\frac{\pi}{9}$ \\[1.0ex]

\phantom{i}$\{3,3,3,2,2,2,2\}$\phantom{i} & 6 &  \large$\frac{\pi}{11}$ \\[1.0ex]
\phantom{i}$\{3,3,3,3,1,1,1\}$\phantom{i} & 6 &  \large$\frac{\pi}{9}$ \\[1.0ex]

\phantom{i}$\{4,4,1,1,1,1,1\}$\phantom{i} & 4 &  \large$\frac{\pi}{9}$ \\[1.0ex]

  \hline\hline
\end{tabular*}
\caption{List of multisets with one or two distinct integers, sums of the equal sum submultisets, and $\chi_{min}$ associated to seven-qubit irreducible $c$-states of maximal length.}
\label{table5}

\end{table}

\begin{table*}[!tb]
\subtable{
\begin{tabular*}{8.6cm}{c c c}
  \hline\hline
  \phantom{i}multiset\phantom{i} & \phantom{i}sum of $P_{+}\backslash{c_{0}}$\phantom{i}& \phantom{i}$\chi_{min}$\phantom{i} \\[1.0ex]
\hline

\phantom{i}$\{3,2,1,1,1,1,1\}$\phantom{i} & 3 &  \large$\frac{\pi}{7}$ \\[1.0ex]

\phantom{i}$\{3,2,2,1,1,1,1\}$\phantom{i} & 3 &  \large$\frac{\pi}{8}$ \\[1.0ex]
\phantom{i}$\{3,2,2,1,1,1,1\}$\phantom{i} & 4 &  \large$\frac{\pi}{7}$ \\[1.0ex]

\phantom{i}$\{3,2,2,2,1,1,1\}$\phantom{i} & 3 &  \large$\frac{\pi}{9}$ \\[1.0ex]
\phantom{i}$\{3,2,2,2,1,1,1\}$\phantom{i} & 4 &  \large$\frac{\pi}{8}$ \\[1.0ex]

\phantom{i}$\{3,2,2,2,2,1,1\}$\phantom{i} & 5 &  \large$\frac{\pi}{8}$ \\[1.0ex]

\phantom{i}$\{3,3,2,1,1,1,1\}$\phantom{i} & 3 &  \large$\frac{\pi}{9}$ \\[1.0ex]
\phantom{i}$\{3,3,2,1,1,1,1\}$\phantom{i} & 4 &  \large$\frac{\pi}{8}$ \\[1.0ex]

\phantom{i}$\{3,3,2,2,1,1,1\}$\phantom{i} & 3 &  \large$\frac{\pi}{10}$ \\[1.0ex]
\phantom{i}$\{3,3,2,2,1,1,1\}$\phantom{i} & 4 &  \large$\frac{\pi}{9}$ \\[1.0ex]
\phantom{i}$\{3,3,2,2,1,1,1\}$\phantom{i} & 5 &  \large$\frac{\pi}{8}$ \\[1.0ex]

\phantom{i}$\{3,3,2,2,2,1,1\}$\phantom{i} & 4 &  \large$\frac{\pi}{10}$ \\[1.0ex]
\phantom{i}$\{3,3,2,2,2,1,1\}$\phantom{i} & 5 &  \large$\frac{\pi}{9}$ \\[1.0ex]

\phantom{i}$\{3,3,2,2,2,2,1\}$\phantom{i} & 6 &  \large$\frac{\pi}{9}$ \\[1.0ex]

\phantom{i}$\{3,3,3,2,1,1,1\}$\phantom{i} & 3 &  \large$\frac{\pi}{11}$ \\[1.0ex]
\phantom{i}$\{3,3,3,2,1,1,1\}$\phantom{i} & 5 &  \large$\frac{\pi}{9}$ \\[1.0ex]

\phantom{i}$\{3,3,3,2,2,1,1\}$\phantom{i} & 4 &  \large$\frac{\pi}{11}$ \\[1.0ex]
\phantom{i}$\{3,3,3,2,2,1,1\}$\phantom{i} & 5 &  \large$\frac{\pi}{10}$ \\[1.0ex]

\phantom{i}$\{3,3,3,2,2,2,1\}$\phantom{i} & 6 &  \large$\frac{\pi}{10}$ \\[1.0ex]

\phantom{i}$\{4,2,2,1,1,1,1\}$\phantom{i} & 4 &  \large$\frac{\pi}{8}$ \\[1.0ex]

\phantom{i}$\{4,2,2,2,1,1,1\}$\phantom{i} & 4 &  \large$\frac{\pi}{9}$ \\[1.0ex]

  \hline\hline
\end{tabular*}}
\subtable{
\begin{tabular*}{8.6cm}{c c c}
  \hline\hline
  \phantom{i}multiset\phantom{i} & \phantom{i}sum of $P_{+}\backslash{c_{0}}$\phantom{i}& \phantom{i}$\chi_{min}$\phantom{i} \\[1.0ex]
\hline

\phantom{i}$\{4,2,2,2,2,1,1\}$\phantom{i} & 4 &  \large$\frac{\pi}{10}$ \\[1.0ex]

\phantom{i}$\{4,3,1,1,1,1,1\}$\phantom{i} & 4 &  \large$\frac{\pi}{8}$ \\[1.0ex]
\phantom{i}$\{4,3,3,1,1,1,1\}$\phantom{i} & 4 &  \large$\frac{\pi}{10}$ \\[1.0ex]
\phantom{i}$\{4,3,3,2,2,2,2\}$\phantom{i} & 6 &  \large$\frac{\pi}{12}$ \\[1.0ex]
\phantom{i}$\{4,3,3,3,2,2,2\}$\phantom{i} & 6 &  \large$\frac{\pi}{13}$ \\[1.0ex]

\phantom{i}$\{4,4,2,1,1,1,1\}$\phantom{i} & 4 &  \large$\frac{\pi}{10}$ \\[1.0ex]
\phantom{i}$\{4,4,2,2,1,1,1\}$\phantom{i} & 4 &  \large$\frac{\pi}{11}$ \\[1.0ex]
\phantom{i}$\{4,4,2,2,1,1,1\}$\phantom{i} & 5 &  \large$\frac{\pi}{10}$ \\[1.0ex]
\phantom{i}$\{4,4,2,2,2,1,1\}$\phantom{i} & 6 &  \large$\frac{\pi}{10}$ \\[1.0ex]
\phantom{i}$\{4,4,2,2,2,1,1\}$\phantom{i} & 4 &  \large$\frac{\pi}{12}$ \\[1.0ex]

\phantom{i}$\{4,4,3,1,1,1,1\}$\phantom{i} & 4 &  \large$\frac{\pi}{11}$ \\[1.0ex]
\phantom{i}$\{4,4,3,3,1,1,1\}$\phantom{i} & 6 &  \large$\frac{\pi}{11}$ \\[1.0ex]

\phantom{i}$\{4,4,3,3,2,2,2\}$\phantom{i} & 6 &  \large$\frac{\pi}{14}$ \\[1.0ex]
\phantom{i}$\{4,4,3,3,2,2,2\}$\phantom{i} & 8 &  \large$\frac{\pi}{12}$ \\[1.0ex]
\phantom{i}$\{4,4,3,3,3,2,2\}$\phantom{i} & 8 &  \large$\frac{\pi}{13}$ \\[1.0ex]

\phantom{i}$\{4,4,4,2,2,1,1\}$\phantom{i} & 6 &  \large$\frac{\pi}{12}$ \\[1.0ex]
\phantom{i}$\{4,4,4,3,3,2,2\}$\phantom{i} & 8 &  \large$\frac{\pi}{14}$ \\[1.0ex]

\phantom{i}$\{5,5,2,2,1,1,1\}$\phantom{i} & 5 &  \large$\frac{\pi}{12}$ \\[1.0ex]
\phantom{i}$\{5,5,2,2,2,1,1\}$\phantom{i} & 6 &  \large$\frac{\pi}{12}$ \\[1.0ex]
\phantom{i}$\{5,5,3,3,1,1,1\}$\phantom{i} & 6 &  \large$\frac{\pi}{13}$ \\[1.0ex]

\phantom{i}$\{6,3,3,3,1,1,1\}$\phantom{i} & 6 &  \large$\frac{\pi}{12}$ \\[1.0ex]
  \hline\hline
\end{tabular*}}

\caption{List of multisets with three distinct integers, sums of the equal sum submultisets, and $\chi_{min}$ associated to seven-qubit irreducible $c$-states of maximal length.}
\label{table6}
\end{table*}

\begin{table*}[!tb]
\subtable{
\begin{tabular*}{8.6cm}{c c c}
  \hline\hline
  \phantom{i}multiset\phantom{i} & \phantom{i}sum of $P_{+}\backslash{c_{0}}$\phantom{i}& \phantom{i}$\chi_{min}$\phantom{i} \\[1.0ex]
\hline

\phantom{i}$\{4,3,2,1,1,1,1\}$\phantom{i} & 4 &  \large$\frac{\pi}{9}$ \\[1.0ex]

\phantom{i}$\{4,3,2,2,1,1,1\}$\phantom{i} & 5 &  \large$\frac{\pi}{9}$ \\[1.0ex]
\phantom{i}$\{4,3,2,2,1,1,1\}$\phantom{i} & 4 &  \large$\frac{\pi}{10}$ \\[1.0ex]

\phantom{i}$\{4,3,2,2,2,1,1\}$\phantom{i} & 5 &  \large$\frac{\pi}{10}$ \\[1.0ex]
\phantom{i}$\{4,3,2,2,2,1,1\}$\phantom{i} & 4 &  \large$\frac{\pi}{11}$ \\[1.0ex]

\phantom{i}$\{4,3,3,2,1,1,1\}$\phantom{i} & 5 &  \large$\frac{\pi}{10}$ \\[1.0ex]

\phantom{i}$\{4,3,3,2,2,1,1\}$\phantom{i} & 6 &  \large$\frac{\pi}{10}$ \\[1.0ex]
\phantom{i}$\{4,3,3,2,2,1,1\}$\phantom{i} & 5 &  \large$\frac{\pi}{11}$ \\[1.0ex]
\phantom{i}$\{4,3,3,2,2,1,1\}$\phantom{i} & 4 &  \large$\frac{\pi}{12}$ \\[1.0ex]

\phantom{i}$\{4,3,3,2,2,2,1\}$\phantom{i} & 6 &  \large$\frac{\pi}{11}$ \\[1.0ex]

\phantom{i}$\{4,3,3,3,2,1,1\}$\phantom{i} & 6 &  \large$\frac{\pi}{11}$ \\[1.0ex]

\phantom{i}$\{4,3,3,3,2,2,1\}$\phantom{i} & 7 &  \large$\frac{\pi}{11}$ \\[1.0ex]

\phantom{i}$\{4,4,3,2,1,1,1\}$\phantom{i} & 5 &  \large$\frac{\pi}{11}$ \\[1.0ex]

\phantom{i}$\{4,4,3,2,2,1,1\}$\phantom{i} & 5 &  \large$\frac{\pi}{12}$ \\[1.0ex]
\phantom{i}$\{4,4,3,2,2,1,1\}$\phantom{i} & 4 &  \large$\frac{\pi}{13}$ \\[1.0ex]
\phantom{i}$\{4,4,3,2,2,1,1\}$\phantom{i} & 6 &  \large$\frac{\pi}{11}$ \\[1.0ex]

\phantom{i}$\{4,4,3,3,2,1,1\}$\phantom{i} & 6 &  \large$\frac{\pi}{12}$ \\[1.0ex]

\phantom{i}$\{4,4,3,3,2,2,1\}$\phantom{i} & 7 &  \large$\frac{\pi}{12}$ \\[1.0ex]

\phantom{i}$\{5,3,2,2,1,1,1\}$\phantom{i} & 5 &  \large$\frac{\pi}{10}$ \\[1.0ex]

\phantom{i}$\{5,3,2,2,2,1,1\}$\phantom{i} & 5 &  \large$\frac{\pi}{11}$ \\[1.0ex]

\phantom{i}$\{5,3,3,2,1,1,1\}$\phantom{i} & 5 &  \large$\frac{\pi}{11}$ \\[1.0ex]

  \hline\hline

\end{tabular*}}
\subtable{
\begin{tabular*}{8.6cm}{c c c}
  \hline\hline
  \phantom{i}multiset\phantom{i} & \phantom{i}sum of $P_{+}\backslash{c_{0}}$\phantom{i}& \phantom{i}$\chi_{min}$\phantom{i} \\[1.0ex]
\hline

\phantom{i}$\{5,3,3,2,2,1,1\}$\phantom{i} & 5 &  \large$\frac{\pi}{12}$ \\[1.0ex]
\phantom{i}$\{5,3,3,2,2,2,1\}$\phantom{i} & 6 &  \large$\frac{\pi}{12}$ \\[1.0ex]

\phantom{i}$\{5,4,2,2,1,1,1\}$\phantom{i} & 5 &  \large$\frac{\pi}{11}$ \\[1.0ex]
\phantom{i}$\{5,4,2,2,2,1,1\}$\phantom{i} & 6 &  \large$\frac{\pi}{11}$ \\[1.0ex]

\phantom{i}$\{5,4,3,3,1,1,1\}$\phantom{i} & 6 &  \large$\frac{\pi}{12}$ \\[1.0ex]

\phantom{i}$\{5,4,4,2,2,1,1\}$\phantom{i} & 6 &  \large$\frac{\pi}{13}$ \\[1.0ex]

\phantom{i}$\{5,4,4,3,3,2,2\}$\phantom{i} & 8 &  \large$\frac{\pi}{15}$ \\[1.0ex]

\phantom{i}$\{5,5,3,2,1,1,1\}$\phantom{i} & 5 &  \large$\frac{\pi}{13}$ \\[1.0ex]

\phantom{i}$\{5,5,3,2,2,1,1\}$\phantom{i} & 5 &  \large$\frac{\pi}{14}$ \\[1.0ex]

\phantom{i}$\{5,5,3,3,2,1,1\}$\phantom{i} & 7 &  \large$\frac{\pi}{13}$ \\[1.0ex]

\phantom{i}$\{5,5,4,2,2,1,1\}$\phantom{i} & 6 &  \large$\frac{\pi}{14}$ \\[1.0ex]

\phantom{i}$\{5,5,4,4,3,3,2\}$\phantom{i} & 10 &  \large$\frac{\pi}{16}$ \\[1.0ex]

\phantom{i}$\{6,3,3,2,2,2,1\}$\phantom{i} & 6 &  \large$\frac{\pi}{13}$ \\[1.0ex]

\phantom{i}$\{6,4,2,2,2,1,1\}$\phantom{i} & 6 &  \large$\frac{\pi}{12}$ \\[1.0ex]

\phantom{i}$\{6,4,3,3,1,1,1\}$\phantom{i} & 6 &  \large$\frac{\pi}{13}$ \\[1.0ex]

\phantom{i}$\{6,4,4,2,2,1,1\}$\phantom{i} & 6 &  \large$\frac{\pi}{14}$ \\[1.0ex]

\phantom{i}$\{6,4,4,3,3,2,2\}$\phantom{i} & 8 &  \large$\frac{\pi}{16}$ \\[1.0ex]

\phantom{i}$\{6,4,4,4,2,1,1\}$\phantom{i} & 8 &  \large$\frac{\pi}{14}$ \\[1.0ex]

\phantom{i}$\{6,5,2,2,2,1,1\}$\phantom{i} & 6 &  \large$\frac{\pi}{13}$ \\[1.0ex]
\phantom{i}$\{6,5,3,3,1,1,1\}$\phantom{i} & 6 &  \large$\frac{\pi}{14}$ \\[1.0ex]

\phantom{i}$\{6,6,4,4,2,1,1\}$\phantom{i} & 8 &  \large$\frac{\pi}{16}$ \\[1.0ex]

  \hline\hline

\end{tabular*}}

\caption{List of multisets with four distinct integers, sums of the equal sum submultisets, and $\chi_{min}$ associated to seven-qubit irreducible $c$-states of maximal length.}
\label{table7}
\end{table*}

\begin{table}[!h]
\begin{tabular*}{8.6cm}{c c c}
  \hline\hline
  \phantom{i}multiset\phantom{i} & \phantom{i}sum of $P_{+}\backslash{c_{0}}$\phantom{i}& \phantom{i}$\chi_{min}$\phantom{i} \\[1.0ex]
\hline
\phantom{i}$\{5,4,3,2,1,1,1\}$\phantom{i} & 5 &  \large$\frac{\pi}{12}$ \\[1.0ex]
\phantom{i}$\{5,4,3,2,2,1,1\}$\phantom{i} & 5 &  \large$\frac{\pi}{13}$ \\[1.0ex]

\phantom{i}$\{5,4,3,2,2,1,1\}$\phantom{i} & 6 &  \large$\frac{\pi}{12}$ \\[1.0ex]
\phantom{i}$\{5,4,3,3,2,1,1\}$\phantom{i} & 6 &  \large$\frac{\pi}{13}$ \\[1.0ex]

\phantom{i}$\{5,4,3,3,2,1,1\}$\phantom{i} & 7 &  \large$\frac{\pi}{12}$ \\[1.0ex]

\phantom{i}$\{5,4,3,3,2,2,1\}$\phantom{i} & 7 &  \large$\frac{\pi}{13}$ \\[1.0ex]

\phantom{i}$\{5,4,4,3,2,1,1\}$\phantom{i} & 7 &  \large$\frac{\pi}{13}$ \\[1.0ex]
\phantom{i}$\{5,4,4,3,2,2,1\}$\phantom{i} & 8 &  \large$\frac{\pi}{13}$ \\[1.0ex]

\phantom{i}$\{5,4,4,3,3,2,1\}$\phantom{i} & 8 &  \large$\frac{\pi}{14}$ \\[1.0ex]

\phantom{i}$\{5,5,4,3,2,1,1\}$\phantom{i} & 7 &  \large$\frac{\pi}{14}$ \\[1.0ex]
\phantom{i}$\{5,5,4,3,2,2,1\}$\phantom{i} & 8 &  \large$\frac{\pi}{14}$ \\[1.0ex]

\phantom{i}$\{6,4,3,2,2,1,1\}$\phantom{i} & 6 &  \large$\frac{\pi}{13}$ \\[1.0ex]
\phantom{i}$\{6,4,3,3,2,1,1\}$\phantom{i} & 6 &  \large$\frac{\pi}{14}$ \\[1.0ex]
\phantom{i}$\{6,4,3,3,2,2,1\}$\phantom{i} & 7 &  \large$\frac{\pi}{14}$ \\[1.0ex]

\phantom{i}$\{6,5,3,3,2,1,1\}$\phantom{i} & 7 &  \large$\frac{\pi}{14}$ \\[1.0ex]
\phantom{i}$\{6,5,4,2,2,1,1\}$\phantom{i} & 6 &  \large$\frac{\pi}{15}$ \\[1.0ex]

\phantom{i}$\{6,5,4,4,2,1,1\}$\phantom{i} & 8 &  \large$\frac{\pi}{15}$ \\[1.0ex]
\phantom{i}$\{6,5,5,4,3,3,1\}$\phantom{i} & 10 &  \large$\frac{\pi}{17}$ \\[1.0ex]

\phantom{i}$\{7,4,3,3,2,2,1\}$\phantom{i} & 7 &  \large$\frac{\pi}{15}$ \\[1.0ex]
\phantom{i}$\{7,5,3,3,2,1,1\}$\phantom{i} & 7 &  \large$\frac{\pi}{15}$ \\[1.0ex]
\phantom{i}$\{7,6,4,4,2,1,1\}$\phantom{i} & 8 &  \large$\frac{\pi}{17}$ \\[1.0ex]

  \hline\hline

\end{tabular*}
\caption{List of multisets with five distinct integers, sums of the equal sum submultisets, and $\chi_{min}$ associated to seven-qubit irreducible $c$-states of maximal length.}
\label{table8}
\end{table}

\begin{table}[!h]
\begin{tabular*}{8.6cm}{c c c}
  \hline\hline
  \phantom{i}multiset\phantom{i} & \phantom{i}sum of $P_{+}\backslash{c_{0}}$\phantom{i}& \phantom{i}$\chi_{min}$\phantom{i} \\[1.0ex]
\hline

\phantom{i}$\{6,5,4,3,2,1,1\}$\phantom{i} & 7 &  \large$\frac{\pi}{15}$ \\[1.0ex]
\phantom{i}$\{6,5,4,3,2,2,1\}$\phantom{i} & 8 &  \large$\frac{\pi}{15}$ \\[1.0ex]
\phantom{i}$\{6,5,4,4,3,2,1\}$\phantom{i} & 9 &  \large$\frac{\pi}{16}$ \\[1.0ex]

\phantom{i}$\{7,5,4,3,2,1,1\}$\phantom{i} & 7 &  \large$\frac{\pi}{16}$ \\[1.0ex]

\phantom{i}$\{7,5,4,3,2,2,1\}$\phantom{i} & 8 &  \large$\frac{\pi}{16}$ \\[1.0ex]

\phantom{i}$\{8,5,4,3,2,2,1\}$\phantom{i} & 8 &  \large$\frac{\pi}{17}$ \\[1.0ex]

  \hline\hline

\end{tabular*}
\caption{List of multisets with six distinct integers, sums of the equal sum submultisets, and $\chi_{min}$ associated to seven-qubit irreducible $c$-states of maximal length.}
\label{table9}
\end{table}

\begin{table}[!h]
\begin{tabular*}{8.6cm}{c c c}
  \hline\hline
 \phantom{i}multiset\phantom{i} & \phantom{i}sum of $P_{+}\backslash{c_{0}}$\phantom{i}& \phantom{i}$\chi_{min}$\phantom{i} \\[1.0ex]
\hline
\phantom{i}$\{7,6,5,4,3,2,1\}$\phantom{i} & 10 &  \large$\frac{\pi}{18}$ \\[1.0ex]

  \hline\hline

\end{tabular*}
\caption{List of multisets with seven distinct integers, sums of the equal sum submultisets, and $\chi_{min}$ associated to seven-qubit irreducible $c$-states of maximal length.}
\label{table10}
\end{table}


\begin{thebibliography}{99}
\bibitem{co}Given a two-qudit state $|\psi\rangle=\sum_{i,j=1}^{d}b_{ij}|{ij}\rangle$, the coefficient matrix is the $d\times{d}$ matrix with elements $b_{ij}$. The determinant of the coefficient matrix is therefore an ${\rm{SU}}(d)\otimes{\rm{SU}}(d)$ invariant polynomial of homogenous degree $d$. For two qubits the determinant of the coefficient matrix is a factor $\frac{1}{2}$ times the complex concurrence.
\end{thebibliography}
\end{document}